# Ultrastrong and ductile CoNiMoAl medium-entropy alloys enabled by L1$_2$ nanoprecipitate-induced multiple deformation mechanisms

Min Young Sung [a], Tae Jin Jang [a], Sang Yoon Song [a], Gunjick Lee [a], KenHee Ryou [b], Sang-Ho Oh [c], Byeong-Joo Lee [c], Pyuck-Pa Choi [b], Jörg Neugebauer [d], Blazej Grabowski [e], Fritz Körmann [d,e,f], Yuji Ikeda [d,e,*], Alireza Zargaran [g,**], Seok Su Sohn [a,***]

[a] *Department of Materials Science and Engineering,*

*Korea University, 02841 Seoul, South Korea*

[b] *Department of Materials Science and Engineering,*

*Korea Advanced Institute of Science and Technology, 34141 Daejeon, South Korea*

[c] *Department of Materials Science and Engineering,*

*Pohang University of Science and Technology, 37673 Pohang, South Korea*

[d] *Computational Materials Design, Max–Planck–Institut für Eisenforschung GmbH,*

*Max–Planck–Straβe 1, 40237 Düsseldorf, Germany*

[e] *Institute for Materials Science,*

*University of Stuttgart, Pfaffenwaldring 55, 70569 Stuttgart, Germany*

[f] *Interdisciplinary Centre for Advanced Materials Simulation (ICAMS),*

*Ruhr-Universität Bochum, 44801 Bochum, Germany*

[g] *Graduate Institute of Ferrous Technology,*

*Pohang University of Science and Technology, 37673 Pohang, South Korea*

**Abstract**

L1$_2$ precipitates are known to significantly enhance the strength and ductility of single-phase face-centered cubic (FCC) medium- or high-entropy alloys (M/HEAs). However, further improvements in mechanical properties remain untapped, as alloy design has historically focused on systems with specific CrCoNi- or FeCoCrNi-based FCC matrix and Ni$_3$Al L1$_2$ phase compositions. This study introduces novel Co-Ni-Mo-Al alloys with L1$_2$ precipitates by systematically altering Al content, aiming to bridge this research gap by revealing the strengthening mechanisms. The (CoNi)$_{81}$Mo$_{12}$Al$_7$ alloy achieves yield strength of 1086 MPa, tensile strength of 1520 MPa, and ductility of 35%, demonstrating an impressive synergy of strength, ductility, and strain-hardening capacity. Dislocation analysis via transmission electron microscopy, supported by generalized stacking fault energy (GSFE) calculations using density functional theory (DFT), demonstrates that Mo substitution for Al in the L1$_2$ phase alters dislocation behavior, promoting the formation of multiple deformation modes, including stacking faults, super-dislocation pairs, Lomer-Cottrell locks, and unusual nano-twin formation even at low strains. These behaviors are facilitated by the low stacking fault energy (SFE) of the FCC matrix, overlapping of SFs, and dislocation dissociation across anti-phase boundaries (APBs). The increased energy barrier for superlattice intrinsic stacking fault (SISF) formation compared to APBs, due to Mo substitution, further influences dislocation activity. This work demonstrates a novel strategy for designing high-performance M/HEAs by expanding the range of FCC matrix and L1$_2$ compositions through precipitation hardening.

\* Corresponding author: Yuji Ikeda (yuji.ikeda@imw.uni-stuttgart.de)

\*\* Corresponding author: Alireza Zargaran (alireza@postech.ac.kr)

\*\*\* Corresponding author: Seok Su Sohn (sssohn@korea.ac.kr)

# 1. Introduction

Medium and high-entropy alloys (M/HEAs) with a single face-centered cubic (FCC) phase have garnered significant attention due to their superior mechanical properties, including strength, ductility, corrosion resistance, and fracture toughness [1,2]. Despite these advantages, the intrinsic yield strength limitations of FCC-based M/HEAs pose a substantial challenge, prompting research into innovative strategies to enhance their mechanical performance. Recent advancements have focused on addressing this challenge through the strategic introduction of uniformly dispersed precipitates within the FCC matrix. This approach aims to obstruct dislocation motion, a key mechanism underlying deformation in metals, thereby enhancing yield strength while maintaining a balance with uniform ductility. Among the various types of explored precipitates, $L1_2$ precipitates have emerged as particularly effective, originating from good coherency with the FCC matrix, resulting in exceptionally high strength and ductility [3,4].

In FCC-based M/HEAs containing $L1_2$ precipitates, the strength and ductility are affected by the characteristics of both $L1_2$ precipitates and FCC matrix [5]. During deformation, $L1_2$ precipitates interact with dislocations, leading to the formation of SFs such as anti-phase boundary (APB), complex stacking fault (CSF), and superlattice intrinsic stacking fault (SISF) [6]. Low APB energy can enhance work hardening and ductility by promoting planar slip [7]. The consecutive CSFs in two layers contribute to pseudo-twin and deformation twin development [8]. SISF induces a local phase transition from the $L1_2$ to the $D0_{19}$ structure, which acts as a thin twin structure [9]. The formation of these faults is associated with the energy of each fault, which is highly dependent on the composition of the $L1_2$ phase [10]. Additionally, the SFE of the matrix also plays a critical role in deformation behaviors. A high SFE of the matrix facilitates the cross-slip of dislocations, leading to the formation of dislocation walls or even microbands, while a low SFE promotes deformation via SFs or twins. Therefore, extensive research has been conducted to improve mechanical properties by varying alloying elements to form $L1_2$ precipitates within the FCC matrix [11]. Nevertheless, further property improvements face untapped potential in the alloy design employing $L1_2$ nano-precipitates strengthening.

The untapped potential is related to the fact that the exploration of L1$_2$ precipitation hardening in M/HEAs has historically been focused on alloy systems with specific FCC matrix and L1$_2$ phase compositions. The compositions of the FCC matrix have been studied in CrCoNi- and FeCoCrNi-based systems [12,13]. In terms of the L1$_2$ phase, Ni$_3$Al, Ni$_3$(Al,Ti), and (Co,Ni)$_3$Al, formed by the addition of Al and Ti in the aforementioned alloy systems, have been studied to enhance the strength [8,12,13]. However, recent advancement has spotlighted CoNiMo [14] alloys for their exceptional mechanical properties, notably their enhanced yield strength, even within a single FCC phase, diverging from traditional FCC single-phase M/HEAs. Through the increased Mo content in CoNiMo alloys, the decrease of SFE of the FCC matrix and the yield strength enhancement facilitate twin formation compared to other conventional M/HEAs. In addition, Mo content also influences the characteristics of L1$_2$ precipitates. According to Zhu et al. [15], the addition of Mo to Ni$_3$Al can enhance the slip barrier of the slip process of three slip systems for the Ni$_3$Al phase, establishing Mo as an excellent strengthening alloying element. However, despite these beneficial effects of Mo, the research landscape remains limited concerning the specific role and potential of L1$_2$ nano-precipitates in strengthening effects within this novel alloy system. Therefore, the strategic introduction of coherent L1$_2$ nano-precipitates offers an opportunity to significantly refine the balance between strength and ductility. In addition, this approach enhances the hardening capacity, which has traditionally been challenging to achieve in the conventional L1$_2$ reinforced M/HEAs.

The present study aims to bridge this research gap by probing into the precipitation strengthening effects, particularly through L1$_2$ nano-precipitates, thereby offering insights into the mechanisms underlying the enhanced mechanical properties of the CoNiMo alloy systems. The formation of L1$_2$ precipitates within a CoNiMo alloy is explored by meticulously adjusting the Al content, guided by thermodynamic predictions. The low SFE of the FCC matrix and excellent strengthening contribution of L1$_2$ precipitates result in superior mechanical properties, including strength, ductility, and strain hardening capability. The primary findings demonstrate that the aged (Co,Ni)$_{81}$Mo$_{12}$Al$_7$ alloy exhibited outstanding properties, with a yield strength of ~1.1 GPa, tensile strength of ~1.5 GPa, and ductility of

~35%. The underlying mechanisms for this exceptional yield strength and strain-hardening capacity were revealed through detailed transmission electron microscopy analyses of multiple deformation modes and dislocation dissociation behaviors, in conjunction with the generalized stacking fault energy (GSFE) obtained via density functional theory (DFT) calculations. The present approach not only holds the potential to improve the combination of strength and ductility in the alloys containing $L1_2$ precipitates but also contributes to broadening the understanding and application scope of precipitation hardening strategies within the field of M/HEAs.

## 2. Experimental methods

### 2.1. Alloy design

As mentioned above, Mo addition in the CoNiMo alloys reduces the SFE and enhances mechanical properties [14]. However, with increased Mo content, as in $(CoNi)_{85}Mo_{15}$ and $(CoNi)_{82}Mo_{18}$, the addition of Al for introducing $L1_2$ phase rather is prone to promote brittle μ phase formation. Therefore, the alloy composition of $(CoNi)_{88}Mo_{12}$ was selected as the base alloy composition. Fig. 1(a) shows the equilibrium pseudo-binary phase diagram of $(CoNi)_{88-x}Mo_{12}Al_x$ (x = 0–10 at.%) alloy system in the temperature range of 400–1600 °C to investigate the role of Al on the variation of the constituent phases in $(CoNi)_{88}Mo_{12}$ MEA. With increasing Al content, the temperature of a single FCC region rises from 800 to 1100 °C. As the temperature decreases, the HCP phase and various intermetallic compounds (IMCs) such as $L1_2$, $Co_3Mo$, and μ phases are predicted in the equilibrium state. The Al composition was selected to prevent the formation of brittle IMC precipitates, such as $Co_3Mo$ and μ, and to achieve a wide-range single FCC region. Therefore, $(CoNi)_{85}Mo_{12}Al_3$ and $(CoNi)_{81}Mo_{12}Al_7$ were carefully chosen as model alloys to investigate the difference in $L1_2$ phase formation with suppressing brittle phases. Two model alloys were produced in dimensions of 100×35×8 mm$^3$ using vacuum induction melting (MC100V, Indutherm, Germany). High-purity elements (>99.9%) were stacked in a zirconia crucible and subsequently cast into a graphite mold under Ar atmosphere. The cast ingots were homogenized at 1200 °C for 2 h in a tube furnace under Ar atmosphere, followed by water-quenching, and then cold-rolled to achieve a thickness reduction of 80%.

To determine the annealing and aging temperatures for both alloys, the phase fractions as a function of temperature (Fig. 1(b,c)) were calculated. A previous study reported that in the $V_{16}(CoNi)_{74}Mo_{10}$ alloy, despite the presence of the µ phase in the equilibrium phase diagram, only a small amount (0.7%) of the µ phase was experimentally observed after heat treatment at 600 °C for 4 h [16]. This result suggests that the formation of the µ phase occurs at a relatively slow rate. Consequently, a short annealing process at 1000 °C for 2 min was used to suppress the formation of µ phase for both the 3Al and 7Al alloys (termed as 3Al and 7Al hereafter), resulting in an FCC single-phase microstructure with refined grains.

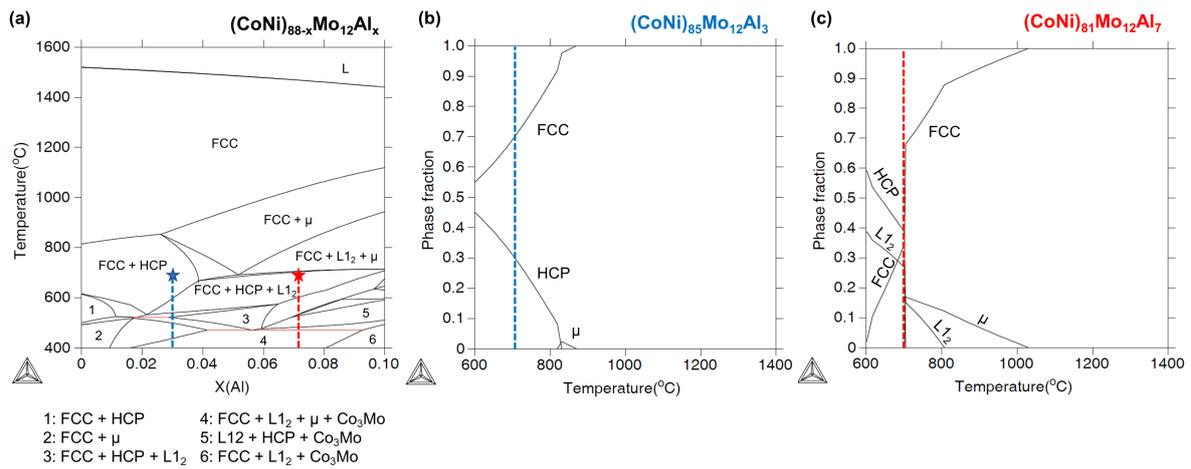

**Fig. 1.** (a) Calculated phase diagram of the $(CoNi)_{88-x}Mo_{12}Al_x$ system as a function of the Al content. Equilibrium phase fractions as a function of temperature for the (b) $(CoNi)_{85}Mo_{12}Al_3$, and (c) $(CoNi)_{81}Mo_{12}Al_7$ alloys.

Following the annealing process, the 3Al and 7Al alloys were aged at 700 °C for 24 h (referred to as 3Al-A and 7Al-A). For the 7Al-A alloy, this aging temperature facilitated an increase in the fraction of the $L1_2$ phase while suppressing µ phase formation (Fig. 1(c)). In the 3Al-A alloy, this temperature led to an HCP phase fraction slightly lower than that of the 7Al-A alloy, without the formation of the $L1_2$ phase. The composition of the HCP and L1$_2$ phases was calculated via thermodynamic calculations, revealing that the HCP phase is Co-rich (Co,Ni)$_3$Mo, while the L1$_2$ phase is Ni-rich (Ni,Co)$_3$Al.

*2.2. Microstructural characterization*

Specimens for metallographic examinations were prepared by grinding with 400, 800, 1200, 2000, and 4000 grit SiC papers, followed by mechanical polishing with 3 and 1 μm diamond suspension. The final polishing was conducted using a colloidal silica suspension for 0.5 h to eliminate any damaged layers on the surface. To identify crystal structures and measure lattice parameters, X-ray diffraction (XRD) analyses with Cu K$_\alpha$ radiation were performed (XRD, D Max 2500, Rigaku, Japan). The scans were carried out from 20° to 100° at a scan rate of 5° min$^{-1}$ with a step size of 0.02°. Grain sizes of the FCC phase after the annealing and aging processes were investigated via electron backscatter diffraction (EBSD, Symmetry S, Oxford Instruments, United Kingdom). The morphology, distribution, and volume fraction of precipitates were investigated via field emission scanning electron microscopy (FE-SEM, Quanta FEG 250, FEI, USA) equipped with a backscattered electron (BSE) detector. Further investigation of the precipitates, deformation structures, and dislocation dissociation behaviors was conducted using transmission electron microscopy (TEM, JEM-2100F, JEOL, Japan) operated at a 200 kV voltage equipped with an EDS detector. The TEM specimens were prepared by grinding to a thickness of 70 μm and then punching into 3 mm diameter disks. These thin foils were subjected to electro-polishing via a twin-jet polisher (Tenupol-5, Struers, Denmark) in a solution composed of 10% perchloric acid and 90% acetic acid with a voltage of 40 V.

Weak beam dark-field (DF) TEM analysis was conducted to calculate the SFE. The g(3g) configuration observed from $\langle 111 \rangle_{FCC}$ zone was used for the specimen tensile-deformed to 5%. Through g·b analysis of three independent $\langle \bar{2}20 \rangle$ diffraction vectors (specifically, g = $(0\bar{2}2)$, g = $(\bar{2}20)$, and g = $(\bar{2}02)$), the Burgers vector of the dislocations was determined. The SFE value was calculated using the following equation [17]:

$$\text{SFE} = \frac{Gb_p^2}{8\pi d_{\text{act}}}\left(\frac{2-\nu}{1-\nu}\right)\left(1 - \frac{2\nu \cos(2\beta)}{2-\nu}\right), \tag{1}$$

where $G$ is the shear modulus, $b_p$ is the magnitude of the Burgers vectors of partial dislocations, $d_{act}$ is the actual dissociation width of the partial dislocations, $\nu$ is Poisson's ratio, and $\beta$ is the angle between the dislocation line and the Burgers vector of perfect dislocations. Cockayne et al. [18] demonstrated that the $d_{act}$ can be calculated by the observed width of partial dislocations. Therefore, the weak beam DF image obtained from the diffraction vector g = ($\bar{2}$02) was utilized to measure the interspacing between the two dissociated partial dislocations, and this value was converted to the value of $d_{act}$ to calculate the SFE.

The detailed chemical composition of the matrix and precipitates was measured by atom probe tomography (APT, LEAP 4000X HR, Cameca Instruments, USA). The laser pulse energy and frequency were 50 pJ and 200 kHz, respectively. The rate of data detection was 0.5% under 50 K, and the APT data were reconstructed and analyzed via the IVAS 3.8.4 software (Cameca Instruments).

*2.3. Mechanical properties test*

To evaluate room-temperature tensile properties, dog-bone-shaped plate tensile specimens with a gauge length of 12.0 mm and width of 4.0 mm were machined along the rolling direction of the annealed and aged alloys. The uniaxial tensile tests were conducted by a universal testing machine (model 3367, Instron, USA) at a strain rate of $10^{-3}$ s$^{-1}$ at room temperature. The tensile strain was measured using an extensometer within the specimen gauge, and the tests were repeated at least three times for reliability. Hardness was measured using a Vickers micro-hardness tester (model MMT-X78, Matsuzawa Co Ltd.) to evaluate the increase in hardness due to the formation of HCP precipitates. Each test was performed with an applied load of 500 gf for 15 s, and ten measurements were carried out for each specimen to attain average hardness values.

*2.4. DFT calculations*

*Ab initio* DFT simulations were conducted to analyze the GSFE of the FCC matrix and the $L1_2$ precipitates in the CoNiMoAl system. $Co_{48}Ni_{48}$, $Co_{42}Ni_{42}Mo_{12}$, and $Co_{38}Ni_{38}Mo_{12}Al_8$ in the FCC phase as well as $Ni_3Al$, $(Co_{1/3}Ni_{2/3})_3Al$, $(Co_{1/3}Ni_{2/3})_3(Al_{3/4}Mo_{1/4})$, and $(Co_{1/3}Ni_{2/3})_3(Al_{1/2}Mo_{1/2})$ in the $L1_2$ phase were considered based on the experimental data to investigate the impact of Mo and Al systematically. The GSFEs of the alloys for the {111} plane were computed by explicitly calculating several selected points on the GSFE surface, followed by a Fourier interpolation [19]. For the explicit calculations, the tilted-supercell approach [19,20] was employed under periodic boundary conditions. For an ideal ordered structure without chemical disorder, rotational, reflectional, and translational symmetries result in sets of equivalent points providing the same GSFE. However, for a disordered alloy model, the points in each set should show different energies. Therefore, to obtain a GSFE reflecting the FCC and the $L1_2$ symmetry, the energies were obtained for each explicit point by averaging over the energies of all symmetrically equivalent points on the considered grid. Further details of the DFT calculation methods are given in Supplemental Materials.

## 3. Results

*3.1. Microstructural evolutions*

X-ray diffraction patterns of the 3Al, 7Al, 3Al-A, and 7Al-A alloys are presented in Fig. 2(a). Only FCC structure peaks are observed in both 3Al and 7Al alloys. Unlike the annealed alloys, the aged alloys show both HCP and $D0_{19}$ peaks in addition to the FCC peaks. A previous study revealed that in CoNiMo alloys, both HCP and $D0_{19}$ phases coexist after heat treatment, contrary to the thermodynamic calculations predicting only the presence of the HCP phase [14]. The absence of $L1_2$ superlattice peaks in the 7Al-A alloy may be due to the peak broadening effect caused by nano-sized precipitates and the small scattering factor due to the complex atomic occupation within the ordered phase [21].

EBSD inverse pole figure maps in Fig. 2(b–e) reveal that fully recrystallized FCC grains are well-developed in all alloys. Both 3Al and 7Al alloys exhibit similar average grain sizes of 3.5 ± 1.9 μm and 3.6 ± 1.5 μm, respectively. These values change negligibly after the aging process, indicating that the formation of $L1_2$ and HCP+$D0_{19}$ precipitates occurs under nearly identical grain sizes. Therefore, this well-controlled grain size facilitates a quantitative investigation of the strengthening effect induced by the precipitates.

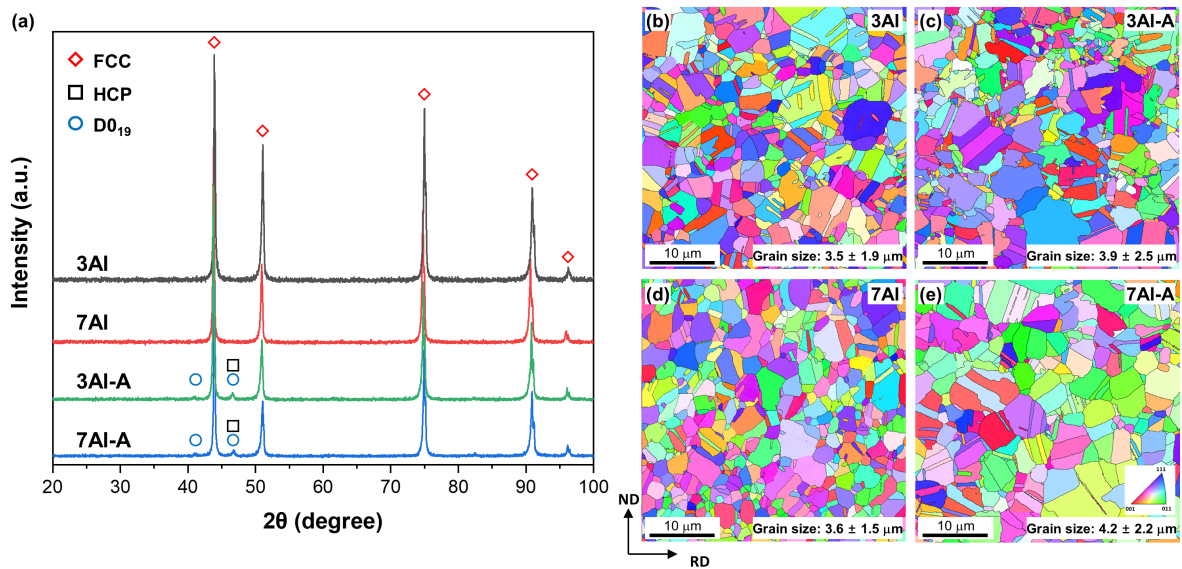

**Fig. 2.** (a) XRD profiles and (b–e) EBSD IPF maps for the 3Al and 7Al alloys annealed at 1000 °C for 2 min, and the 3Al-A and 7Al-A alloys aged at 700 °C for 24 h. Red, black, and blue symbols in (a) indicate the FCC, HCP, and $D0_{19}$ phases, respectively.

Fig. 3(a–d) shows SEM-BSE images of the 3Al, 7Al, 3Al-A, and 7Al-A alloys. Both annealed alloys show a recrystallized FCC structure with annealing twins (Fig. 3(a,c)), confirming the absence of precipitates, consistent with XRD results. In contrast, the 3Al-A and 7Al-A alloys exhibit plate-like precipitates distributed uniformly along grain boundaries and at triple junctions. Considering the XRD results and thermodynamic calculations, these precipitates can be expected to possess the HCP+$D0_{19}$ structure. The volume fractions of these precipitates in the 3Al-A and 7Al-A alloys are 3.7% and 2.7%, indicating a slightly higher fraction in the 3Al-A alloy.

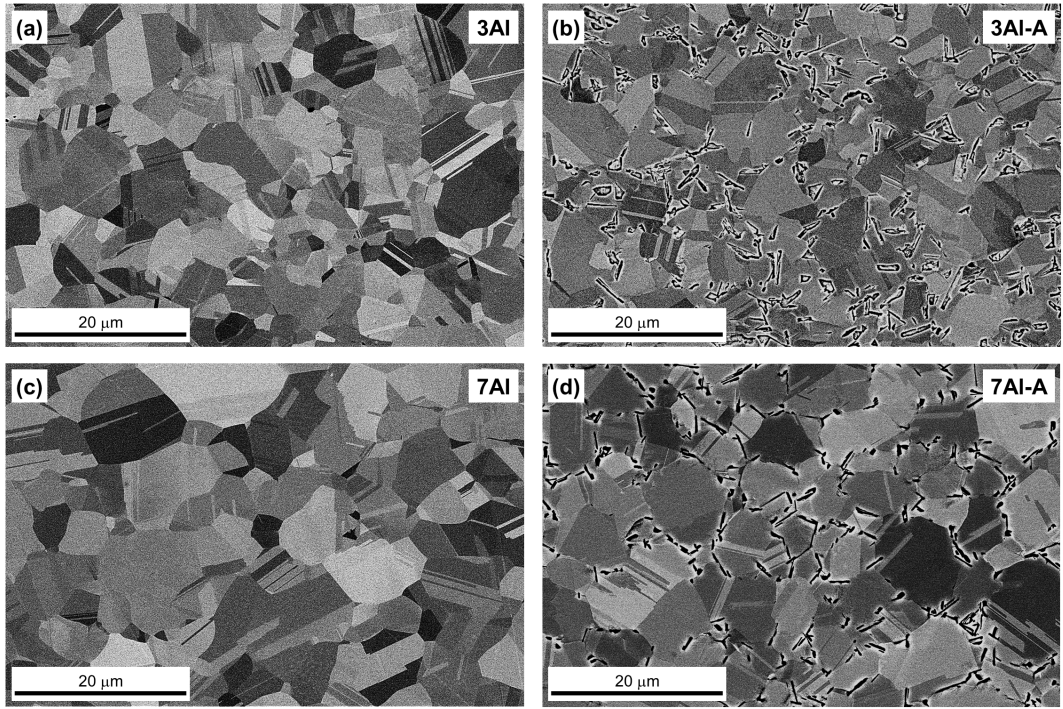

**Fig. 3.** SEM-BSE images of the (a) 3Al, (b) 3Al-A, (c) 7Al, and (d) 7Al-A alloys, showing FCC matrix with annealing twins and plate-like precipitates along grain boundaries and triple junctions.

TEM analysis was conducted to unravel the specific crystal structures and detailed chemical compositions for the aged alloys. For the 3Al-A alloy, Fig. 4(a–d) reveals that the precipitates at grain boundaries exhibit a layered structure composed of the FCC and HCP+$D0_{19}$ phases in a plate-like manner. The TEM-DF images obtained from the $D0_{19}$ and FCC spot (Fig. 4(b,c)) show the distinct contrast difference along the directions indicated by arrows (e) and (f), respectively. The HR-TEM image and the Fast Fourier Transformation (FFT) analysis (Fig. 4(d)) confirm that the contrast difference along arrow (e) is attributed to the layered structure of the HCP and $D0_{19}$ phases, while the contrast difference along arrow (f) is related to the FCC phase. To further clarify the layered structure of the FCC and HCP+$D0_{19}$ phases and the localized phase transformation between the HCP and $D0_{19}$ within the precipitates, TEM-EDS was conducted to determine the compositional difference between the HCP, $D0_{19}$, and FCC phases. Fig. 4(e) shows no compositional differences between the HCP and $D0_{19}$ phases at 45Co-31Ni-23Mo-1Al (at.%). This result corresponds to the previous report that the formation of the $D0_{19}$ phase can be attributed to a partial disorder-order transformation, considering the stability of Co$_3$Mo with the D0$_{19}$ structure at low temperatures and the same compositions between the HCP and D0$_{19}$ phases [14]. On the other hand, in Fig. 4(f), the FCC phase within the precipitates exhibits higher Ni and Al contents and lower Co and Mo contents compared to the HCP+D0$_{19}$ phases. Thus, the composition of FCC and HCP+D0$_{19}$ phases are 40Co-45Ni-10Mo-5Al and 45Co-31Ni-23Mo-1Al (at.%), respectively.

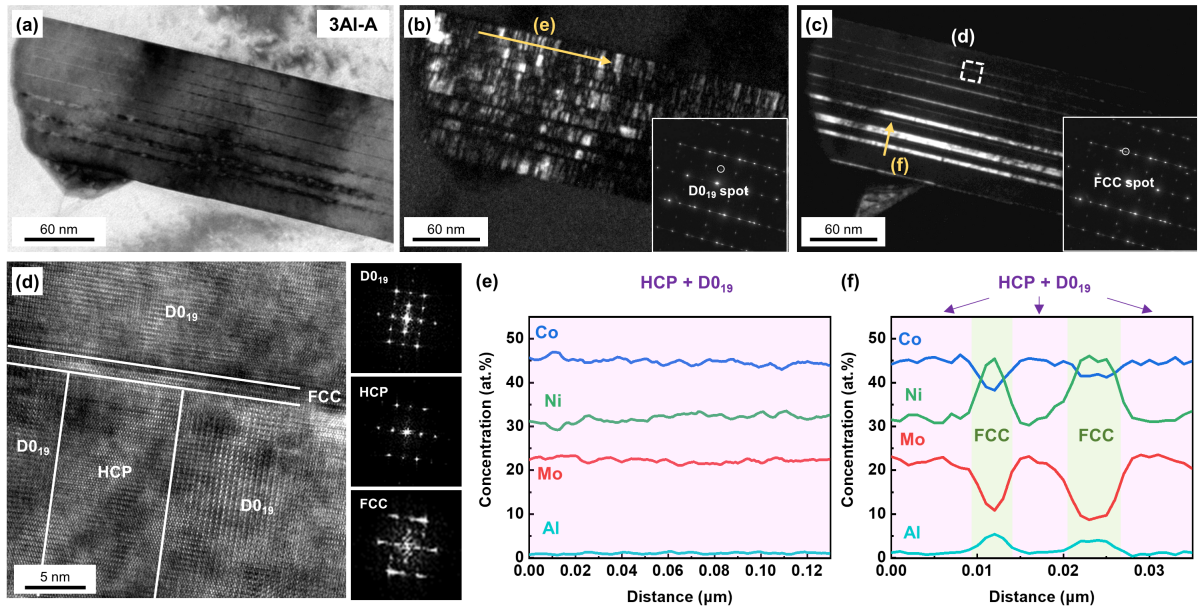

**Fig. 4.** (a) TEM-BF and (b,c) corresponding DF images of grain-boundary precipitates in the 3Al-A alloy with SADP patterns for the (b) HCP + D0$_{19}$ layers and (c) FCC layers. (d) HR-TEM images of the white box in (c) showing the FCC layers and HCP + D0$_{19}$ layers, with FFT patterns of each phase. TEM-EDS line profile along the orange arrow of (e) and (f) shown in (b) and (c), respectively.

Such compositional differences of Mo and Al in FCC and HCP+D0$_{19}$ phases locally affect the stability of the FCC, HCP, and D0$_{19}$ phases. Feng et al. [22] demonstrated that as the Mo content decreases and the Al content increases in the Co-Al-W-Ta-B-Mo alloy system, the FCC phase becomes more stable than the Co$_3$Mo of D0$_{19}$ structure. Therefore, a localized decrease in Mo content and an increase in Al content may promote the formation of the FCC structure, leading to a layered structure between the FCC and HCP+D0$_{19}$ phases. Fig. S1(a) provides representative scanning TEM (STEM) images at low magnification for the 3Al-A alloy, revealing the formation of the HCP+D0$_{19}$ precipitates at the FCC grain boundaries, consistent with the SEM observation (Fig. 3(b)). The selected area diffraction pattern (SADP) along the ⟨110⟩$_{FCC}$ zone axis confirms the coexistence of HCP and D0$_{19}$ phases within the precipitates and the absence of L1$_2$ precipitates (Fig. S1(a,b)). The FCC and HCP+D0$_{19}$ phases exhibit a crystallographic orientation relationship known as the Shoji-Nishiyama (S-N) relation, with ⟨11$\bar{2}$0⟩$_{HCP}$ //⟨110⟩$_{FCC}$ and (0002)$_{HCP}$//(111)$_{FCC}$ [23]. Thus, the 3Al-A alloy consists of the FCC matrix, along with the precipitates of the layered HCP, D0$_{19}$, and FCC phases.

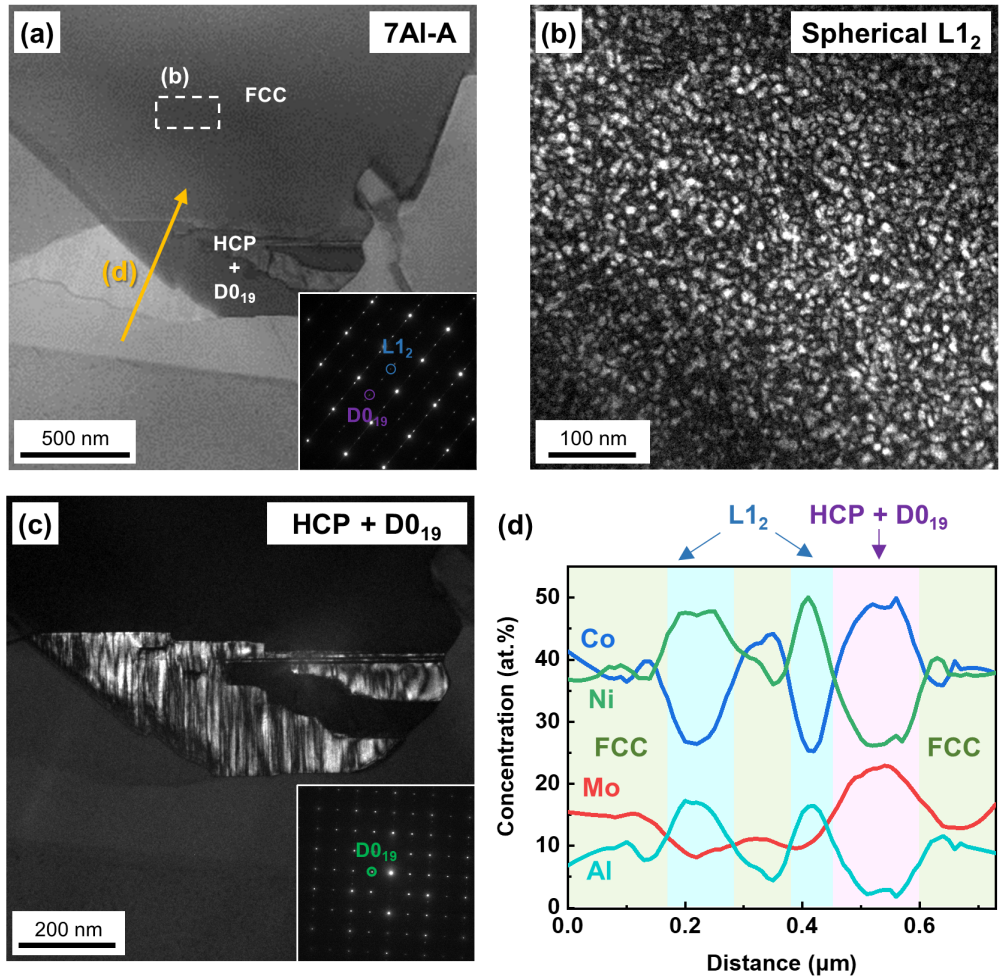

**Fig. 5.** (a) STEM image with the corresponding SADPs showing FCC matrix, L1$_2$ precipitate, and HCP + D0$_{19}$ precipitate in 7Al-A alloy. DF images of (b) L1$_2$ precipitates in FCC matrix, corresponding to the regions indicated in (a). (c) DF image of HCP + D0$_{19}$ precipitate in (a). (d) TEM-EDS line profiles along the orange arrow shown in (a).

For the 7Al-A alloy, Fig. 5(a) shows the STEM image with SADP along the ⟨110⟩$_{FCC}$ zone axis, revealing the coexistence of the FCC matrix, HCP+D0$_{19}$ precipitates, and L1$_2$ precipitates. The L1$_2$ precipitates have a near-spherical shape, with a radius of 6.14 nm and a volume fraction of 37.6% (Fig. 5(b)). On the other hand, the HCP+D0$_{19}$ precipitates in the 7Al-A alloy (Fig. 5(c)) exhibit vertical and horizontal contrast differences and morphology similar to those observed in the 3Al-A alloy (Fig. 4(b,c)). TEM-EDS analysis (Fig. 5(d)) reveals that the HCP+D0$_{19}$ phase has a composition similar to that of 3Al-A, consisting of 49Co-26Ni-23Mo-2Al (at.%), while the L1$_2$ phase exhibits a composition of 26Co-49Ni-8Mo-17Al (at.%).

APT analysis was performed to investigate elemental partitioning within each phase at the atomic scale in both 3Al-A and 7Al-A alloys (Fig. 6(a–d)). The three-dimensional reconstruction map of the 3Al-A alloy (Fig. 6(a)) exhibits Mo-enriched regions while not revealing Al-rich regions. The TEM-EDS and thermodynamic calculations indicate that the L1$_2$ phase is enriched in Ni and Al, while the HCP+D0$_{19}$ phase is rich in Mo. Thus, the APT result implies the presence of the HCP+D0$_{19}$ phase but the absence of the L1$_2$ phase in the 3Al-A alloy. Fig. 6(b) illustrates the compositions of the FCC matrix and Mo-rich regions along the indicated arrow, with compositions of 47Co-38Ni-12Mo-3Al (at.%) and 50Co-29Ni-20Mo-1Al (at.%), respectively.

For the 7Al-A alloy, the three-dimensional reconstruction map (Fig. 6(c)) shows the presence of both Al-rich and Mo-rich regions, with compositions of 31Co-45Ni-11Mo-13Al (at.%) and 46Co-32Ni-17Mo-5Al (at.%), respectively (Fig. 6(d,e)). The FCC matrix displays a composition of 48Co-37Ni-11Mo-4Al (at.%), which is comparable to that of the 3Al-A. The 1-D concentration profiles reveal that the L1$_2$ precipitates exhibit a similar composition regardless of the presence of surrounding Mo-rich regions. The compositions of the L1$_2$ and HCP+D0$_{19}$ precipitates of the 7Al-A alloy slightly deviate from the thermodynamic calculations (Table 1). Specifically, the L1$_2$ phase exhibits higher Co and Mo contents, with a rise of 13 and 7 at.%, respectively, while the HCP+D0$_{19}$ phase shows a higher content of Ni by 8 at.% and Al by 5 at.%. However, the sum of Co and Ni contents and that of Mo and Al contents in the L1$_2$ phase are approximately 76 at.% and 24 at.%, respectively, closely corresponding to each sum of contents based on the thermodynamic calculations. This result suggests a mutual substitution of Co with Ni and that of Mo with Al during a simultaneous formation of the HCP+$D0_{19}$ and $L1_2$ precipitates in the 7Al-A alloy. When considering the FCC matrix composition in 7Al-A alloy and utilizing the lever rule with this value, the volume fraction of the $L1_2$ precipitates derived from the APT data is 31±3%, closely approximating the value of 37.6% observed in TEM. Regarding the morphology of the $L1_2$ precipitates, the precipitates appear to be interconnected by necks. This morphology arises from the synergistic effects of their small size and high volume fraction, resulting in a reduced edge-to-edge inter-precipitate distance [24,25].

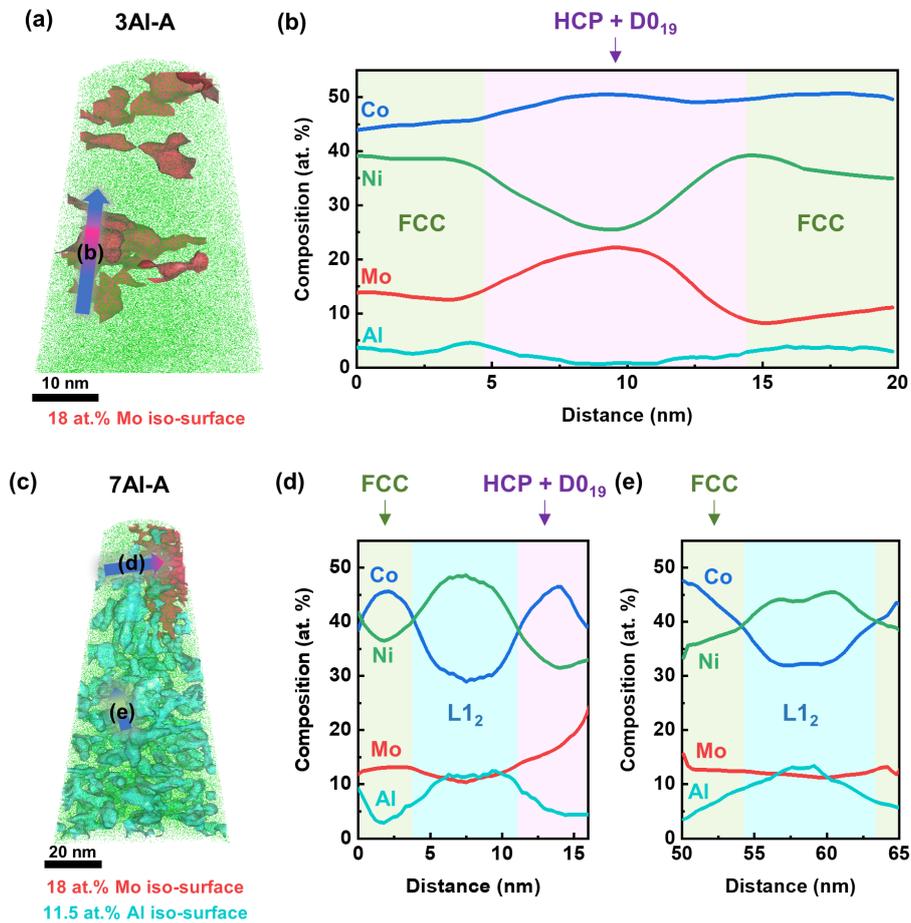

**Fig. 6.** APT elemental distribution analysis of the 3Al-A and 7Al-A alloys. (a) Three-dimensional reconstruction map for the 3Al-A alloy and (b) one-dimensional compositional profiles of each element along the arrow shown in (a). (c) Three-dimensional map for the 7Al-A alloy and (d,e) one-dimensional compositional profiles of each element along the arrow shown in (c).

*3.2. GSFE calculations*

The APT data of the 7Al-A alloy (Fig. 6(d,e)) reveals that, in comparison to the Ni$_3$Al composition, approximately 41% of Ni sites are substituted with Co atoms, and approximately 46% of Al sites are substituted with Mo atoms in the L1$_2$ precipitates. To analyze the impact of these substitutions systematically, GSFEs were calculated through DFT calculations for the L1$_2$ phases with a composition of Ni$_3$Al as well as (Co$_{1/3}$Ni$_{2/3}$)$_3$Al, (Co$_{1/3}$Ni$_{2/3}$)$_3$(Al$_{3/4}$Mo$_{1/4}$), and (Co$_{1/3}$Ni$_{2/3}$)$_3$(Al$_{1/2}$Mo$_{1/2}$), where Co and Mo partially substitute the Ni and the Al sublattices, respectively. Generally, the presence of L1$_2$ precipitates leads to the formation of various types of SFs due to their interaction with dislocations. The glide of $\frac{a}{6}\langle\bar{1}\bar{1}2\rangle$ partial dislocation forms CSF, $\frac{a}{2}\langle\bar{1}10\rangle$ partial dislocation glide results in the formation of the APB, and $\frac{a}{3}\langle11\bar{2}\rangle$ partial dislocation glide leads to the formation of SISF. We therefore particularly focus on the values of stable and unstable SFEs for the CSF, the APB, and the SISF states.

Table 2 summarizes the obtained stable and unstable SFEs of the investigated L1$_2$ phases, and Fig. 7(a,b) shows the GSFE curves of Ni$_3$Al and (Co$_{1/3}$Ni$_{2/3}$)$_3$(Al$_{1/2}$Mo$_{1/2}$), respectively, for the path from the no-fault (NF) state to the CSF, APB, and SISF states. (The GSFE curves for the other L1$_2$ phases are shown in Fig. S8.) $\gamma_i$ denotes the value of SFE of each fault. Compared to Ni$_3$Al, (Co$_{1/3}$Ni$_{2/3}$)$_3$(Al$_{1/2}$Mo$_{1/2}$) shows the increase of the stable SFE values for the CSF and APB ($\gamma_{CSF}$ and $\gamma_{APB}$) by 189 and 161 mJ/m$^2$, respectively, and a decrease of the stable SISF energy ($\gamma_{SISF}$) by 44 mJ/m$^2$. In the case of the unstable SFE, the unstable SISF energy ($\gamma_{USISF}$) exhibits the highest increase of 195 mJ/m$^2$, followed by the unstable CSF energy ($\gamma_{UCSF}$) of 169 mJ/m$^2$ and the unstable APB energy ($\gamma_{UAPB}$) of 146 mJ/m$^2$. These significant changes in SFEs are expected to impact dislocation dissociation behavior in the 7Al-A alloy, which will be further analyzed in the subsequent Section 4.2.

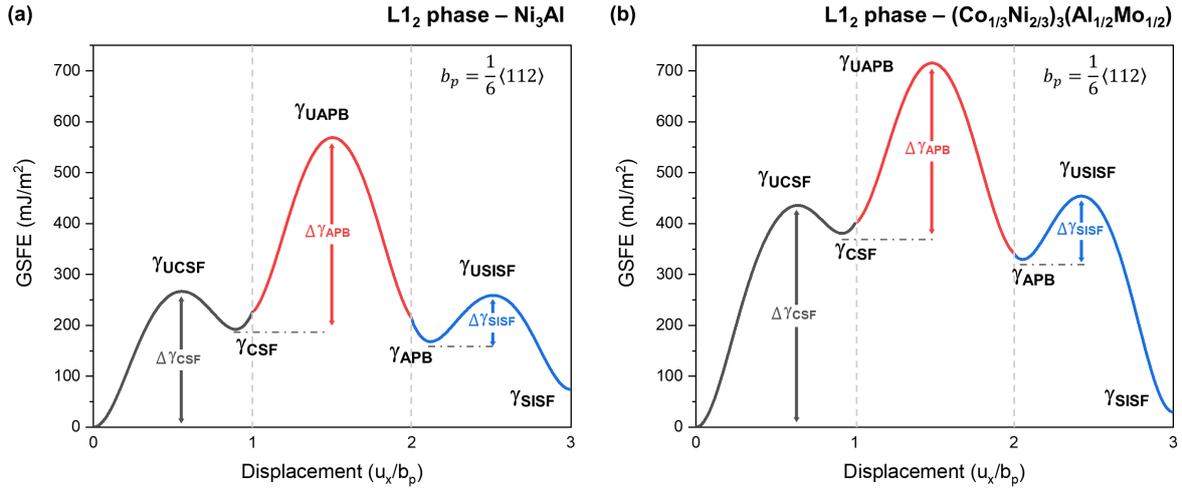

**Fig. 7.** GSFE curves of (a) the L1$_2$-Ni$_3$Al and (b) the L1$_2$-(Co$_{1/3}$Ni$_{2/3}$)$_3$(Al$_{1/2}$Mo$_{1/2}$) along the NF-CSF-APB-SISF path.

*3.3. Mechanical properties*

Fig. 8(a) shows the engineering stress-strain curves of the four alloys and their mechanical properties are summarized in Table 3. The 3Al and 7Al alloys exhibit similar mechanical properties with yield strength (YS) of 652 MPa and 668 MPa, tensile strength (TS) of 1088 MPa and 1127 MPa, and elongation (El.) of 59% and 62%, respectively. After the aging process, the 3Al-A alloy shows a marginal increment of YS by 112 MPa, while the 7Al-A alloy exhibits a YS of 1086 MPa, significantly increasing by 418 MPa. Similarly, the increase of tensile strength is more significant in the 7Al-A alloy exhibiting 1520 MPa than the 3Al-A alloy of 1202 MPa. In contrast to the increase in strength, ductility is reduced to 48% and 35% for the 3Al-A and 7Al-A alloys, respectively.

Fig. 8(b,c) reveals the strain-hardening rate (SHR) curves of the four alloys. The 3Al and 3Al-A alloys exhibit similar trends of a monotonic decline of SHR, regardless of the heat treatment. Nonetheless, the 3Al-A alloy shows a higher hardening rate than the 3Al alloy up to around 14% true strain, attributed to the presence of the HCP+D0$_{19}$ precipitates at grain boundaries. Conversely, unlike the 3Al alloys, the SHR curves of the 7Al alloy show significant changes by the aging process. The 7Al alloy exhibits a comparable trend and value to the 3Al alloy, showing a gradual decrease as the strain increases. In contrast, the 7Al-A alloy reveals superior strain-hardening capability with a rise up to 10%

strain with a peak rate of 4800 MPa. This outstanding strain hardening in the 7Al-A alloy is attributed to the presence of both hard HCP+$D0_{19}$ precipitates and $L1_2$ precipitates. In particular, the interactions between the dislocations and $L1_2$ precipitates result in the formation of various planar defects, such as super-dislocation pairs, APBs, and CSFs, contributing to the enhancement of strain-hardening capacity, which will be elucidated in the following section.

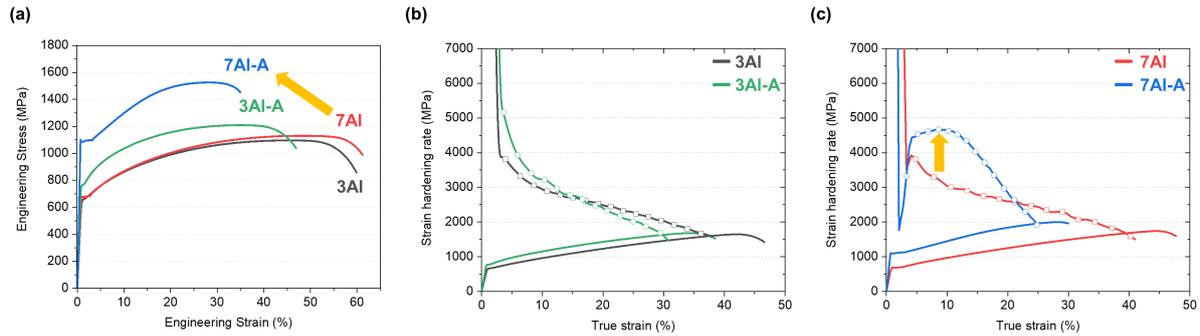

**Fig. 8.** (a) Room-temperature engineering stress-strain curves of all alloys, (b) strain-hardening rate curves for the 3Al and 3Al-A alloys, and (c) strain-hardening rate curves for the 7Al and 7Al-A alloys.

## 4. Discussion

*4.1. Evolution of deformation substructures*

Section 3.3 highlighted the distinct SHR behaviors of the 3Al and 7Al alloys post-aging, particularly noting the SHR increase at early deformation stages for the 7Al-A alloy. To further examine the mechanisms, TEM analysis was performed on the 3Al-A and 7Al-A alloys, tensile-deformed to low (~5%) and high (~20%) true strains, based on the peak hardening rate at 10% true strain as a reference. This section aims to clarify the microstructural differences driving the observed SHR variations.

*4.1.1. 5% true strain*

Fig. 9($a_1$–$a_4$) shows the deformation substructure of the 3Al-A alloy at 5% true strain. In the overall view of the deformation substructure (Fig. 9($a_1$)), a network of dislocations and SFs is arranged along various {111} planes and accumulates at annealing twin boundaries. Fig. 9($a_2$) demonstrates that the

3Al-A alloy deformed through SF and dislocation slip, and the corresponding SADP confirms no evidence of twinning reflections or mechanical micro-twins. Further analysis using the HR-TEM image and FFT analysis (Fig. 9($a_3$)) validates the absence of nano-twinning, indicating that SFs are the primary deformation feature without the presence of twin spots [26].

To analyze the deformation behavior of the 3Al-A alloy, its SFE value was determined using TEM weak beam analysis. The SFE of the FCC matrix, as derived from TEM data, is approximately 18.8 mJ/m$^2$. In FCC alloys, typically, dislocation slip becomes the dominant deformation mechanism for SFE values higher than 45 mJ/m$^2$, whereas twinning is more active when SFE lies between 15 to 45 mJ/m$^2$ [27]. For the 3Al-A alloy, despite the low SFE of 18.8 mJ/m$^2$, evident twin spots are not observed. This difficulty in twin formation can be attributed to the high critical stress required for twin formation. This stress is affected by several factors, including grain size and the SFE itself.

The critical twinning stress ($\sigma_t$) was calculated using the following equation [28]:

$$\sigma_t = \frac{m\gamma}{b_p} + \frac{k_t}{D^{1/2}} \qquad (2)$$

where $m$ is the Taylor factor ($m$ = 3.06) [29], $\gamma$ is the SFE ($\gamma$ = 18.8 mJ/m$^2$), $b_p$ is the magnitude of the Burgers vector of a partial dislocation ($b_p = \frac{\sqrt{6}}{6}a$), $a$ is the lattice constant of the FCC matrix, $D$ is the average grain size, and $k_t$ is the Hall-Petch coefficient for twinning. For FCC alloys, $k_t = 2 \times k_S$ is approximately estimated based on the correlations between $k_t$ and $k_s$ [30,31], where 949 MPa·μm$^{1/2}$ of $k_S$ is the Hall-Petch coefficient for dislocation slip (Fig. S2(a)). The engineering stress-strain curves for the 3Al and 7Al alloys, subjected to different annealing temperatures resulting in various grain sizes, are shown in Fig. S2(b,c). According to these calculations, the contribution from the SFE and Hall-Petch coefficient is 394 MPa and 817 MPa, respectively, resulting in an overall critical twinning stress of 1211 MPa. This value, approximately 24% higher than the flow stress at 5% true strain for the 3Al-A alloy (975 MPa), suggests that the flow stress at this strain level is insufficient to initiate twin formation. The calculated result is consistent with the absence of twin spots in the TEM data (Fig. 9($a_2$,$a_3$)).

In the deformation microstructures of the 7Al-A alloy (Fig. 9($b_1$–$b_4$)), dislocations and SFs accumulate at annealing twin boundaries and align along various {111} planes. The STEM image (Fig. 9($b_2$)), reveals numerous SFs, super-dislocation pairs, and Lomer-Cottrell locks. The super-dislocations, consisting of dislocation pairs with identical Burgers vectors gliding along the same slip plane, serve to minimize the energy arising from the interaction between dislocations and the L1$_2$-ordered structure. In this process, the initial unit dislocation with $\frac{a}{2}\langle 110 \rangle$ Burgers vector glides through the L1$_2$-ordered structure, generating APBs. To eliminate these APBs created during the glide, the subsequent dislocation with the same Burgers vector also glides through the same slip plane [32].

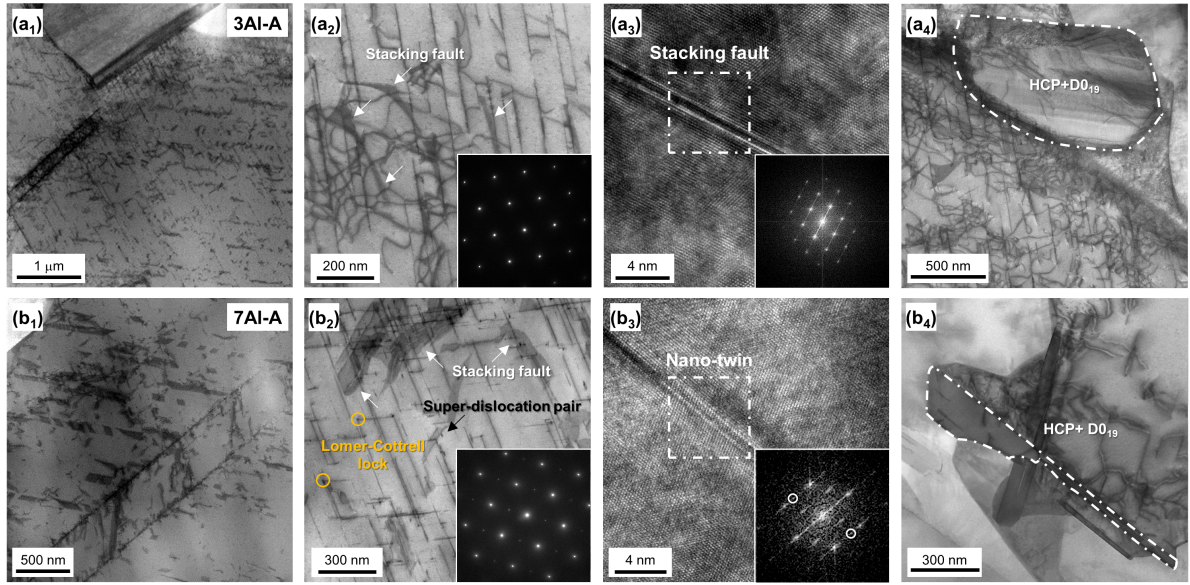

**Fig. 9.** Deformed microstructure of the 3Al-A and 7Al-A alloys deformed to 5% true strain. STEM images of ($a_1$, $a_2$) 3Al-A and ($b_1$, $b_2$) 7Al-A alloys showing dislocations and SFs developed along {111} planes. HR-TEM image and FFT pattern of the white box region of ($a_3$) 3Al-A and ($b_3$) 7Al-A alloys, showing twin spots and SFs. STEM image of ($a_4$) 3Al-A and ($b_4$) 7Al-A alloy, exhibiting dislocation pile up at the interface between FCC matrix and HCP+D0$_{19}$ precipitate.

The L1$_2$ precipitates in the 7Al-A alloy exhibit a high value of 329 mJ/m$^2$ in $\gamma_{APB}$ due to the substitution of Mo for Al, as shown in Fig. 7(b). This increased APB energy implies that APBs, formed through interactions with dislocations, should disappear to achieve a more stable configuration. Therefore, it can be inferred that super-dislocation pairs exist from a 5% true strain, aimed at mitigating the APBs induced by the interaction between $L1_2$ precipitates and dislocations. The super-dislocation pairs and SFs facilitate the formation of SF networks and induce severe dislocation entanglement. This interaction among the defects renders dislocation glide more difficult, thereby enhancing hardening capacity. Based on the corresponding SADP (Fig. 9($b_2$)), the absence of twin spots suggests that micro-twins are not formed, similar to the 3Al-A alloy. However, HR-TEM and FFT analysis (Fig. 9($b_3$)) reveal both twin spots and SFs, indicating the emergence of nano-twins within the 7Al-A alloy even at a 5% strain. This twin formation is uncommon in conventional FCC-$L1_2$ structured alloys, where twin nucleation is limited by the high SFE of the $L1_2$ structure and spatial constraints for twin nucleation [7,33]. The occurrence of nano-twins is attributed to three critical factors, as further detailed below: the continuous overlapping of SFs along {111} planes; the low SFE of the matrix in the 7Al-A alloy; and the substitution of Mo for Al in the $L1_2$ precipitates.

The overlapping of SFs can be attributed to the presence of numerous continuous SFs along the {111} plane, as observed in the TEM results (Fig. 9($b_1$)). These continuous SFs enhance the possibility of SF interactions, implying that planes with different atomic arrangements can exist continuously. These planes with altered arrangements increase the potential for nano-twin formation. Consequently, the alignment of SFs and subsequent overlapping contribute to the formation of nano-twins [34,35].

Another factor influencing nano-twin formation is the low SFE of the matrix. A low SFE indicates comparable stability between the FCC and HCP stacking sequences within the matrix. This similarity promotes extensive faulting as the dominant deformation mechanism, thereby increasing the frequency of fault generation and, consequently, the probability of twin nucleation [36]. Based on the APT data (Fig. 6), the FCC matrix compositions of 3Al-A and 7Al-A are nearly identical, implying a similar SFE for the FCC matrix in 7Al-A. Additionally, the 7Al-A alloy exhibits a significantly high flow stress of 1225 MPa at 5% true strain due to the formation of $L1_2$ precipitates. These two factors synergistically promote twin formation in the FCC matrix. However, the overlapping of SFs and the low SFE of the matrix have been also observed in conventional alloys with FCC and L1$_2$ microstructures. Therefore, unlike the Ni$_3$Al-L1$_2$ in conventional alloys, the L1$_2$ composition of (Co$_{1/3}$Ni$_{2/3}$)$_3$(Al$_{1/2}$Mo$_{1/2}$) emerges as a crucial factor in the generation of nano-twins, affecting $\gamma_{UCSF}$ and pseudo-twin formation.

In a microstructure of FCC matrix and L1$_2$ precipitate, the CSF formation affects twin formation. The glide of a single Shockley partial dislocation with $\frac{a}{6}\langle\bar{1}\bar{1}2\rangle$ leads to CSF formation by interacting with L1$_2$ precipitates. Subsequent gliding of dislocations with the same Burgers vector on adjacent planes generates two CSFs, creating highly unstable Al-Al nearest neighbor bond pairs. At high temperatures, atom reordering eliminates these unstable Al-Al bonds, resulting in the development of a complete twin. However, at relatively lower temperatures where atom diffusion is limited, reordering is difficult to occur, leading to the emergence of pseudo twins [8], which facilitate nano-twin formation by thickening through the shear of $\frac{a}{6}\langle\bar{1}\bar{1}2\rangle$ partial dislocation on consecutive planes [37].

TEM-EDS (Fig. 5(d)) and APT (Fig. 6(d,e)) results indicate that the L1$_2$ precipitates in the 7Al-A alloy are enriched by Mo, deviating from the Ni$_3$Al composition. This compositional change leads to the lowest value for $\gamma_{UCSF}$ compared to $\gamma_{UAPB}$ and $\gamma_{USISF}$ (Table 2), indicating the reduction of the CSF formation barrier. In Ni$_3$Al-L1$_2$, the mentioned two CSFs on consecutive {111} planes form a pseudo-twin by altering the L1$_2$ to an orthorhombic structure, which creates high-energy Al-Al bonds [38,39]. However, when Al in L1$_2$ is substituted with Mo, the probability of the formation of these high-energy Al-Al bonds is reduced [40]. Additionally, when Al is fully replaced by Mo, the ground state of Ni$_3$Mo is the orthorhombic D0$_a$ structure, which has a stacking sequence similar to the orthorhombic-based crystal formed by two stacked CSFs. Thus, the creation of successive CSFs under the Mo replacement favors the formation of pseudo-twins. Consequently, the interaction between the L1$_2$ precipitates and dislocations leads to the formation of SFs, super-dislocation pairs, Lomer-Cottrell locks, and nano-twins from the early stages of deformation. These deformation mechanisms result in a notably high hardening rate at the onset of deformation, contributing to a peak hardening rate of 4800 MPa.

Accordingly, the 3Al-A and 7Al-A alloys demonstrate distinct deformation behaviors, particularly in terms of nano-twin formation. Despite these differences, both alloys exhibit similar deformation substructures regarding the HCP+D0$_{19}$ precipitates. These precipitates exhibit a higher hardness value compared to the relatively soft FCC matrix, which originates from the intrinsic property of IMCs. The hardness difference causes dislocations to pile up at the interface between the FCC matrix and HCP+D0$_{19}$ precipitate (Fig. 9(a$_4$,b$_4$)), implying that these precipitates are not easily sheared by gliding dislocations and serve as barriers to dislocation movement. Thus, the accumulation of dislocations generates back stress, enhancing the flow stress and resulting in a higher SHR compared to conditions where such precipitates are absent [41].

*4.1.2. 20% true strain*

Fig. 10(a$_1$–a$_3$) illustrates the deformation microstructures in the 3Al-A alloy at 20% true strain. Fig. 10(a$_1$) shows that dislocation glide and SFs remain the predominant deformation mechanism at 20% true strain, consistent with observations at 5% true strain. However, the HR-TEM image and FFT analysis reveal the emergence of nano-twins at this higher strain level (Fig. 10(a$_2$)). This occurrence of nano-twins is supported by the calculated critical twinning stress. The flow stress at 20% true strain was recorded at 1434 MPa, which surpasses the calculated critical twinning stress of 1211 MPa. Thus, it is evident that the 3Al-A alloy achieves the requisite stress to facilitate twin formation at 20% strain.

In the 7Al-A alloy (Fig. 10(b$_1$)), there is a notable increase in SF formation. These SFs form an SF network along various {111} planes, and the spacing between the network decreases as its density increases, thereby impeding dislocation glide and resulting in a higher SHR [42]. Furthermore, the extension and overlapping of SFs are observed, as shown in Fig. 10(b$_1$). The extended and overlapped SFs interact with each other and create the Lomer-Cottrell locks. These locks effectively block the dislocation movements, leading to an increase in the SHR [43]. The HR-TEM image and FFT analysis also reveal the presence of nano-twins as well as the SFs (Fig. 10(b$_2$)), indicating a complex interplay of deformation mechanisms.

Unlike the matrix deformation, as the strain increases up to 20%, the HCP+D0$_{19}$ precipitates in both alloys exhibit similar deformation microstructures (Fig. 10(a$_3$,b$_3$)). Despite the increased strain, the hard HCP+D0$_{19}$ precipitates exhibit limited deformation, with notable dislocation pile-up occurring at the interface between the HCP+D0$_{19}$ precipitates and the FCC matrix. The TEM images in Fig. 10(a$_3$,b$_3$) confirm that the HCP+D0$_{19}$ precipitates act as barriers to dislocation movement. However, the SHR value for the 7Al-A alloy is approximately 500 MPa higher than that of the 7Al alloy, unlike the 3Al and 3Al-A alloys exhibiting similar SHR values. This discrepancy underlines the more pronounced impact of L1$_2$ precipitates on enhancing the SHR value compared to the influence exerted by HCP+D0$_{19}$ precipitates.

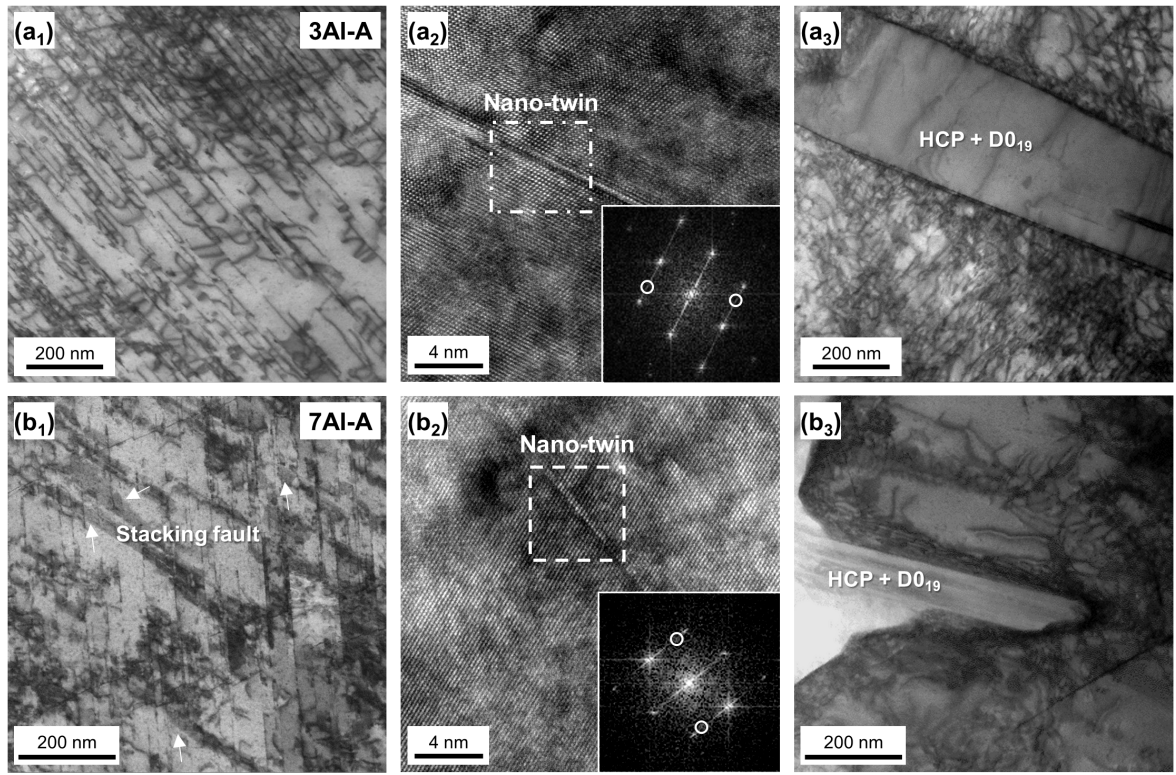

**Fig. 10.** Deformed microstructure of the 3Al-A and 7Al-A alloys deformed to 20% true strain. STEM image of (a$_1$) 3Al-A and (b$_1$) 7Al-A alloy, showing well-developed dislocations and SFs network. HR-TEM image and FFT pattern of white box region of (a$_2$) 3Al-A and (b$_2$) 7Al-A alloys showing atomic image and nano-twin formations. STEM images of (a$_3$) 3Al-A and (b$_3$) 7Al-A alloys, showing dislocation pile up at the interface between FCC matrix and HCP + D0$_{19}$ precipitate.

In summary, the deformation mechanisms of the 3Al-A and 7Al-A alloys exhibit distinct characteristics depending on the strain amounts. Initially, the 3Al-A alloy undergoes deformation through dislocation slip and SFs, as the flow stress remains below the threshold for critical twinning stress. As strain progresses, twin formation becomes increasingly evident. Conversely, from the onset, the 7Al-A alloy activates multiple deformation modes (e.g., SFs, super-dislocation, Lomer-Cottrell locks, and nano-twins), leading to a marked rise in SHR. This behavior persists through higher strains, providing enhanced strain hardening capacity. The dislocation motion resulting from the interaction between $L1_2$ precipitates and dislocations is closely associated with these deformation behaviors. Thus, understanding the origin behind the distinct dislocation dissociation behaviors in the 3Al-A and 7Al-A alloys is critical for comprehending their mechanical response under strain and will be further explored in Section 4.2.

*4.2. Dislocation dissociation behaviors*

For the 3Al-A alloy without $L1_2$ precipitates, the SFE of the FCC matrix was determined to be a low value of 18.8 mJ/m$^2$, as obtained through weak beam DF analysis. This low SFE implies that the $\frac{a}{2}[1\bar{1}0]$ perfect dislocations can easily dissociate into partial dislocations, such as $\frac{a}{6}[1\bar{2}1] + \frac{a}{6}[2\bar{1}\bar{1}]$ dislocations, as shown in the dislocation dissociation analysis at 5% true strain (Fig. 11(a$_1$,a$_2$)).

On the other hand, in the 7Al-A alloy, the presence of $L1_2$ precipitates is the critical factor that alters the dislocation behaviors due to their interactions with dislocations. Typically, when $L1_2$ precipitates exist, dislocations on the $(11\bar{1})$ plane exhibit two different dislocation dissociation mechanisms, as follows [44]:

$$2 \times \frac{a}{2}[1\bar{1}0] \rightarrow \frac{a}{6}[2\bar{1}1] + CSF + \frac{a}{6}[1\bar{2}\bar{1}] + APB + \frac{a}{6}[2\bar{1}1] + CSF + \frac{a}{6}[1\bar{2}\bar{1}] \tag{3}$$

$$2 \times \frac{a}{2}[1\bar{1}0] \rightarrow \frac{a}{6}[2\bar{1}1] + CSF + \frac{a}{6}[1\bar{2}\bar{1}] + APB + \frac{a}{6}[112] + SISF + \frac{a}{6}[\bar{1}\bar{1}\bar{2}] + APB +$$
$$\frac{a}{6}[2\bar{1}1] + CSF + \frac{a}{6}[1\bar{2}\bar{1}] \tag{4}$$

As shown in Eqs. 3 and 4, the interactions between the L1$_2$ precipitates and dislocations lead to various SFs, such as CSF, ABP, and SISF. The stability of these SFs, determined by their respective stable and unstable SFE values, directly affects the favorable SF types and dislocation dissociation behaviors. In the 7Al-A alloy, the presence of super-dislocation pairs and their further dissociation are shown in Fig. 11(b$_1$,b$_2$). Each super-dislocation, with a Burgers vector of $\frac{a}{2}[1\bar{1}0]$, dissociates into two partial dislocations, initiating SF sequences as described in Eq. (3). The leading partial dislocation of the leading super-dislocation glides through the L1$_2$ precipitate, generating a CSF. Subsequently, the trailing partial dislocation of the leading super-dislocation gliding transforms the CSF into an APB. Due to the high APB energy, the leading partial dislocation of the trailing super-dislocation eliminates the APB. Consequently, the leading partial dislocation of the trailing super-dislocation changes the APB back to CSF, and then the glide of trailing partial dislocation of the trailing super-dislocation annihilates the CSF.

This dislocation dissociation behavior is influenced by not only the unstable and stable SFE values but also the energy barrier associated with dislocation motion depending on the SF sequence. In DFT calculations (Fig. 7(a) and Table 2), the unstable and stable SFEs of Ni$_3$Al follow the order of $\gamma_{UAPB}$ > $\gamma_{UCSF}$ > $\gamma_{USISF}$ and $\gamma_{CSF}$ > $\gamma_{APB}$ > $\gamma_{SISF}$, respectively. The L1$_2$ structure of (Co$_{1/3}$Ni$_{2/3}$)$_3$(Al$_{1/2}$Mo$_{1/2}$) exhibits the same order in the stable SFE values as Ni$_3$Al, but the order of unstable SFE values differs. Due to the significant increase in the $\gamma_{USISF}$, it becomes greater than $\gamma_{UCSF}$, resulting in the order $\gamma_{UAPB}$ > $\gamma_{USISF}$ > $\gamma_{UCSF}$. This suggests that the formation of SISF is more difficult, while the formation of CSF becomes easier in the L1$_2$ precipitates of (Co$_{1/3}$Ni$_{2/3}$)$_3$(Al$_{1/2}$Mo$_{1/2}$).

According to Eqs. (3) and (4), whether a dislocation dissociates into a CSF or SISF when an APB is already formed, is determined by the energy barriers encountered in each sequence. Chen et al. [45] explained that the forces on the leading and trailing partial dislocations, depending on the SF sequence, are calculated based on the differences in stable energy values of each SF, and the sign of these differences is crucial. A negative energy difference indicates a thermodynamically spontaneous entry, while a positive value means that stress is required for each dislocation to move. In the case of $(Co_{1/3}Ni_{2/3})_3(Al_{1/2}Mo_{1/2})$, each sequence in Eq. (3) shows that transitioning from APB to CSF has $\gamma_{CSF} - \gamma_{APB}$ of +52 mJ/m². In contrast, transitioning from the CSF to the NF state has $\gamma_{NF} - \gamma_{CSF}$ of -381 mJ/m², indicating that the stress on the leading partial dislocation is less than 100 mJ/m² and the partial dislocation moves into the $L1_2$ precipitates thermodynamically spontaneously. Regarding each sequence in Eq. (4), transitioning from APB to SISF results in a highly negative $\gamma_{SISF} - \gamma_{APB}$ of -299 mJ/m² for the leading partial dislocation, while changing from SISF back to APB for the trailing partial requires the opposite positive value. This positive value is much higher than the value of $\gamma_{CSF} - \gamma_{APB}$ in Eq. (3), preventing the dissociation sequence of Eq. (4) in $(Co_{1/3}Ni_{2/3})_3(Al_{1/2}Mo_{1/2})$.

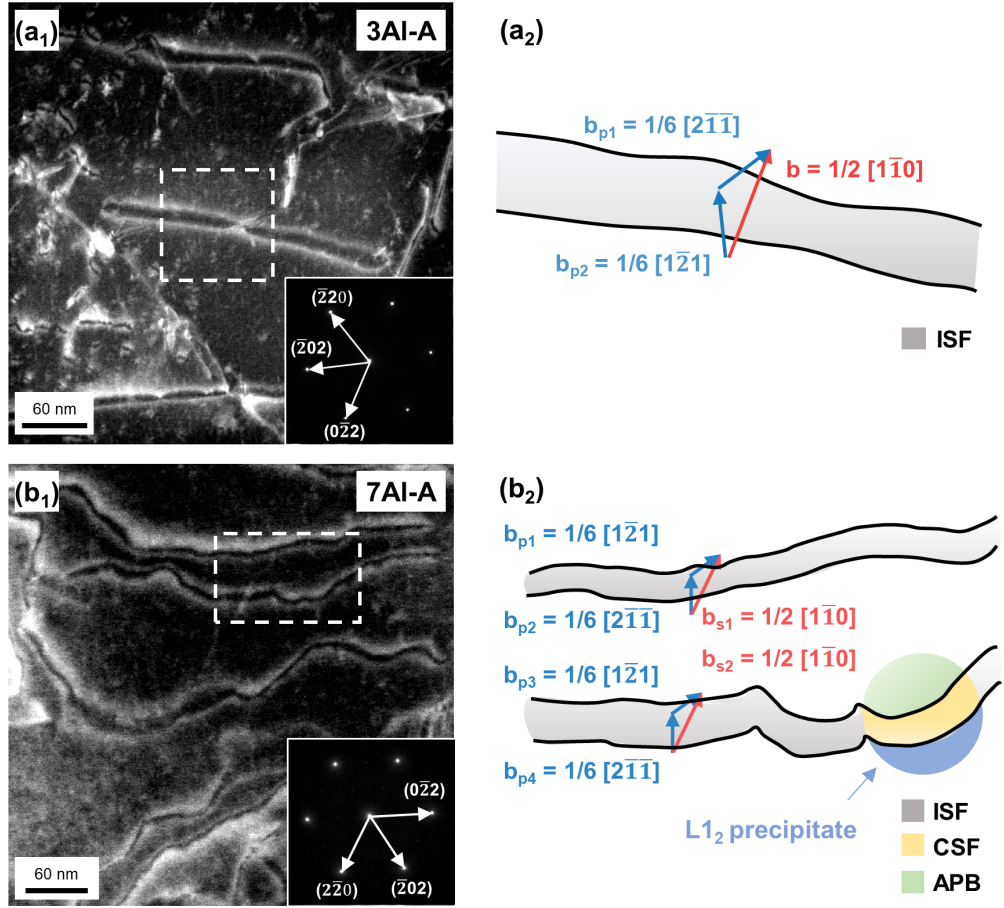

**Fig. 11.** Dislocation dissociation analysis of 3Al-A and 7Al-A alloys deformed to 5% true strain. Weak beam TEM-DF images of (a₁) 3Al-A and (b₁) 7Al-A alloys using g = $\bar{2}20$, g = $\bar{2}02$, and g = $0\bar{2}2$,

respectively. Schematic diagram of dislocation dissociation and each type of planar fault in (a$_2$) 3Al-A and (b$_2$) 7Al-A alloys.

Therefore, the L1$_2$ composition of (Co$_{1/3}$Ni$_{2/3}$)$_3$(Al$_{1/2}$Mo$_{1/2}$) facilitates the formation of CSFs and APBs according to the sequence in Eq. (3), but makes SISF formation more challenging in the sequence in Eq. (4) for the 7Al-A alloy. The presence of these SFs from early-stage deformation facilitates the increase in SF density, the formation of SF networks, and the formation of nano-twins, which in turn significantly contributes to the remarkable tensile strength of 1520 MPa. This remarkable tensile strength results from not only a high hardening capacity but also a high YS of 1087 MPa. Therefore, the following section will focus on explaining the influence of L1$_2$ and HCP+D0$_{19}$ precipitates on achieving the high YS.

*4.3. Strengthening mechanisms*

As revealed in Fig. 8(a), the increment of YS ($\Delta\sigma_{0.2}$) for the 3Al-A alloy is 112 MPa, whereas the 7Al-A alloy experiences a more substantial increase of 418 MPa upon aging, resulting in an outstanding YS of 1087 MPa. This result highlights the predominant influence of L1$_2$ precipitates over HCP+D0$_{19}$ precipitates in enhancing the YS. Therefore, in this section, a quantitative analysis will be conducted regarding the various hardening effects contributing to the YS increase, particularly focusing on the precipitates, to provide a deeper understanding of the hardening mechanisms.

In polycrystalline materials, the enhancement of YS is typically attributed to several mechanisms: solid-solution hardening ($\Delta\sigma_s$), dislocation hardening, grain-boundary hardening ($\Delta\sigma_{gb}$), and precipitation hardening ($\Delta\sigma_P$) [46]. Given that all alloys in this study exhibit a fully recrystallized microstructure, the contribution of dislocation hardening is considered negligible. By subtracting the YS of the 3Al and 7Al alloys from that of the 3Al-A and 7Al-A alloys, the contributions of solid-solution hardening and grain-boundary hardening can be minimized. This approach assumes the grain sizes across all compared alloys are consistent, thereby attributing YS improvement predominantly to precipitation hardening. Thus, the YS increase of the 3Al-A and 7Al-A alloys primarily reflects the sum of their respective contributions, as follows:

$$\Delta\sigma_{0.2} = \Delta\sigma_{p(L1_2)} + \Delta\sigma_{p(HCP+D0_{19})} \tag{5}$$

In the case of the 3Al-A alloy, the absence of $L1_2$ precipitates means that the observed YS increment of 112 MPa primarily stems from the presence of HCP+$D0_{19}$ precipitates. This increment can be further substantiated by leveraging the established correlation between hardness and YS found in various studies [47,48]. The present FCC matrix of the 3Al alloy exhibits a hardness of 2.79 GPa, while the hardness of $Co_3(Mo,Al)$ with a $D0_{19}$ structure is notably higher at 11.6 GPa [49]. As the aging process progresses, the fraction of the FCC matrix decreases in favor of HCP+$D0_{19}$ precipitate formation. Consequently, the formation of $Co_3(Mo,Al)$ precipitates effectively increases the overall hardness, which is calculated by subtracting the FCC matrix hardness from the precipitate hardness and then adjusting for the volume fraction of the precipitates. The increase in hardness is measured to be 325 MPa using the following equation:

$$\Delta H_{V\,HCP+D019} = (H_{V\,Co_3(Mo,Al)} - H_{V\,FCC\,matrix}) \times V_{Co_3(Mo,Al)} \tag{6}$$

where $H_V$ is the Vickers hardness value and $V$ is the volume fraction. Tian et al. [50] demonstrated that the hardness and YS of FCC-based HEAs follow the empirical relationship shown in the equation:

$$\sigma_{0.2} = 0.38 \times H_V - 370 \tag{7}$$

Therefore, the YS increase due to the hardness increase can be obtained by multiplying the hardness increase by 0.38. With the hardness increase quantified at 325 MPa for the 3Al-A alloy relative to the 3Al alloy, this formula yields a YS increment of approximately 125 MPa. This calculation closely aligns with the observed YS difference of 112 MPa between the 3Al and 3Al-A alloys, indicating that the YS improvement in the 3Al-A alloy by the aging process is predominantly due to the formation of HCP+$D0_{19}$ precipitates.

In contrast to the 3Al-A alloy, both HCP+D0$_{19}$ and L1$_2$ precipitates form during the aging process for the 7Al-A alloy. Firstly, the YS increment due to the HCP+D0$_{19}$ precipitates was calculated using the same methodology applied to the 3Al-A alloy. Considering the matrix hardness of the 7Al alloy (2.91 GPa) and the volume fraction of the HCP+D0$_{19}$ precipitates in the 7Al-A alloy (2.7%), the calculated increments in hardness and YS due to the HCP+D0$_{19}$ precipitates amounts to 235 MPa and 89.4 MPa, respectively. In both 3Al-A and 7Al-A alloys, despite the high hardness of the HCP+D0$_{19}$ precipitates, their low volume fraction results in a limited increase in strength.

For the L1$_2$ precipitates in the 7Al-A alloy, the strengthening effect is determined by either one of two mechanisms: Orowan bowing or dislocation shearing, depending on which one is easier to activate. The Orowan bowing mechanism occurs when precipitates are large enough to prevent dislocations from passing through the L1$_2$ precipitates, resulting in dislocation pinning around the precipitates. On the other hand, the dislocation shearing mechanism comes into play when precipitates are small enough to allow dislocations to cut through them [51]. In the 7Al-A alloy, due to the small radius of the L1$_2$ precipitates of 6.14 nm, hardening attributable to these precipitates is predominantly assessed through the dislocation shearing mechanism. Within the particle shearing mechanism, three factors contribute to the increase in YS: coherency strengthening ($\Delta\sigma_{cs}$), modulus mismatch strengthening ($\Delta\sigma_{ms}$), and order strengthening ($\Delta\sigma_{os}$). The total strength enhancement due to dislocation shearing is represented by the larger value between $\Delta\sigma_{cs} + \Delta\sigma_{ms}$ and $\Delta\sigma_{os}$ [51]. The equations for calculating these contributions are as follows [52, 53]:

$$\Delta\sigma_{cs} = M\alpha_\varepsilon (G\varepsilon_c)^{\frac{3}{2}} \left(\frac{rf}{0.5Gb}\right)^{\frac{1}{2}} \qquad (8)$$

$$\Delta\sigma_{ms} = M 0.0055 (\Delta G)^{\frac{3}{2}} \left(\frac{2f}{G}\right)^{\frac{1}{2}} \left(\frac{r}{b}\right)^{\frac{3m}{2}-1} \qquad (9)$$

$$\Delta\sigma_{os} = M 0.81 \frac{\gamma_{APB}}{2b} \left(\frac{3\pi f}{8}\right)^{\frac{1}{2}} \qquad (10)$$

where $M$=3.06 is the Taylor factor, $\alpha_\varepsilon$=2.6 for FCC materials [52], $m$=0.85 is a constant [53], and $G$=92.4 GPa is the shear modulus of (CoNi)$_{88}$Mo$_{12}$ at room temperature, determined via linear fitting from a previous study [14]. $\varepsilon_c$ is the constrained lattice parameter misfit, $2/3(\Delta a/a)$, denoting the lattice parameter mismatch between the matrix and the precipitates as identified by HR-TEM analysis. $r$ is the mean precipitate radius, $f$ is the volume fraction of the precipitates, $b$ is the magnitude of the matrix Burgers vector, $\Delta G$ is the shear modulus mismatch between the matrix and the precipitates, and $\gamma_{APB}$ is the APB energy of precipitates ($\Delta G$=92.4–77=15.4 GPa, $\gamma_{APB}$=0.12 J/m$^2$ are adopted form the corresponding data of Ni$_3$Al precipitates in Ni-based superalloys [54]).

Applying these parameters, $\Delta\sigma_{os}$ yields a higher value of 391 MPa compared to the combined value of $\Delta\sigma_{cs}$ (144 MPa) + $\Delta\sigma_{ms}$ (220 MPa), which totals 364 MPa. This result indicates that order strengthening is the predominant mechanism contributing to YS in the 7Al-A alloy. Therefore, for the 7Al-A alloy, the YS is contributed by solid-solution hardening (463 MPa) and grain-boundary hardening (147 MPa), and further enhanced by the presence of HCP+D0$_{19}$ precipitates (89 MPa) and the L1$_2$ precipitates (391 MPa). These cumulative contributions result in a total calculated yield strength of 1090 MPa, which closely aligns with the experimental value of 1086 MPa, demonstrating the significant role of L1$_2$ precipitates alongside other hardening mechanisms in the enhanced mechanical performance.

Fig. 12 presents an Ashby plot, which compares the tensile strength with elongation of the present alloy against those of conventional M/HEAs containing various precipitates within the FCC matrix. These include FCC+L1$_2$ [4,55–58], FCC+L1$_2$+B2 [59–61], FCC+L1$_2$+L2$_1$ [62,63], FCC+B2 [64–67], and FCC+σ [68], μ [69–71], and M$_{23}$C$_6$ [72,73]. Here, the 7Al-A alloy stands out due to its exceptional combination of strength, ductility, and hardening capacity, characterized by a tensile strength exceeding 1.5 GPa and an elongation surpassing 35%. This achievement highlights its superior performance over conventional FCC-based M/HEAs.

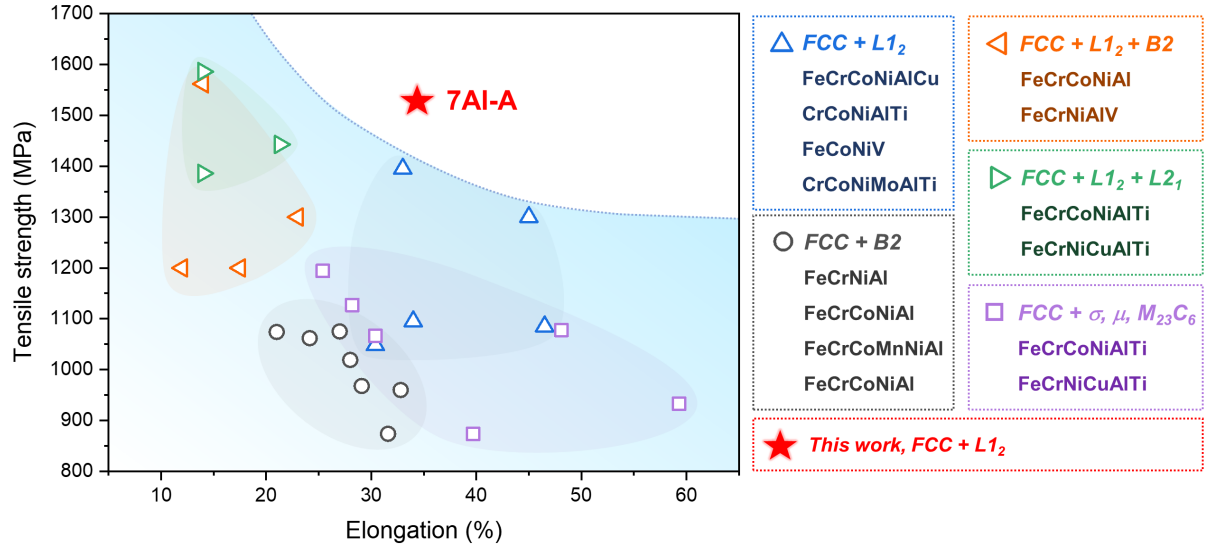

**Fig. 12.** Ashby plot showing the tensile strength versus the elongation for the present alloy in comparison with those of FCC+L1$_2$[4,55–58] M/HEAs, FCC+L1$_2$+B2[59–61] M/HEAs, FCC+L1$_2$+L2$_1$[62,63] M/HEAs, FCC+B2[64–67] M/HEAs, and FCC+σ[68], μ[69–71], M$_{23}$C$_6$[72,73] M/HEAs.

## 5. Conclusions

This investigation delves into the roles of L1$_2$ precipitates in the (CoNi)$_{88-x}$Mo$_{12}$Al$_x$ (x = 3,7 at.%) alloy system, focusing on their mechanical properties, deformation mechanisms, and dislocation dissociation behaviors. A comprehensive analysis based on DFT calculations and empirical assessment has provided key insights as summarized in the following:

(1) The addition of Al enabled the formation of L1$_2$ phase in the CoNiMo alloy system. Proper selection of the 3Al and 7Al compositions has prevented brittle IMC formation, achieving comparable microstructures of the FCC matrix with the HCP phase while differing in L1$_2$ presence. The 3Al-A alloy comprised 3.7% HCP+D0$_{19}$ precipitates within the FCC matrix, whereas the 7Al-A alloy contained 2.7% HCP+D0$_{19}$ and a significant fraction of 37.6% of L1$_2$ precipitates with a radius of 6.14 nm in the FCC matrix.

(2) Aging induced a YS increase of 112 MPa in the 3Al-A alloy, contrasting sharply with a 418 MPa increase in the 7Al-A alloy, resulting in an outstanding YS of 1086 MPa for the latter. Although the 3Al-A alloys exhibited a monotonic decline of the SHR, the 7Al-A alloy revealed a superior strain-hardening capability, with a peak SHR of 4800 MPa, significantly enhancing its tensile strength to 1520 MPa.

(3) For the 3Al-A alloy, dislocation slip and SF were deformation processes at low strains, but nano-twin formation became apparent at high strain levels due to the high critical twinning stress. In contrast, the 7Al-A alloy exhibited various deformation modes, including SFs, super-dislocation, Lomer-Cottrell locks, and nano-twins at low strains. Especially, the formation of nano-twin was attributed to the low SFE of the matrix, SF overlapping, and notably Al substitution with Mo in $L1_2$ precipitates.

(4) DFT calculations on GSFE revealed that in the $L1_2$ phase, Co and Mo substitution for Ni and Al changed the unstable and stable SFE values, increasing the energy barrier for SISF and decreasing it for the APB, which promoted dislocation dissociation through APB rather than SISF. These SFE changes affected the early-stage formation of SFs and nano-twins observed in the 7Al-A alloy.

(5) The YS increment by 120 MPa for the 3Al-A alloys was attributed to HCP+$D0_{19}$ precipitates, whereas for the 7Al-A alloy, HCP+$D0_{19}$ precipitates and $L1_2$ precipitates contributed to the YS increase of 89 MPa and 391 MPa, respectively. Hence, the formation of $L1_2$ precipitates not only markedly enhanced the YS but also activated various deformation modes, achieving a remarkable tensile strength of 1520 MPa and 35% ductility.

**CRediT authorship contribution statement**

**Min Young Sung**: Conceptualization, Methodology, Validation, Formal analysis, Investigation, Writing - Original Draft, Writing - Review & Editing; **Tae Jin Jang**: Validation, Writing - Review &

Editing; **Sang Yoon Song**: Validation, Writing - Review & Editing; **Gunjick Lee**: Methodology, Validation; **KenHee Ryou**: Methodology, Validation; **Sang-Ho Oh**: Investigation; **Byeong-Joo Lee**: Supervision, Validation; **Pyuck-Pa Choi**: Supervision, Validation; **Jörg Neugebauer**: Supervision, Validation; **Blazej Grabowski**: Supervision, Validation; **Fritz Körmann**: Supervision, Validation, Writing - Review & Editing; **Yuji Ikeda**: Supervision, Validation, Writing - Review & Editing; **Alireza Zargaran**: Investigation, Visualization, Supervision, Validation, Writing - Review & Editing; **Seok Su Sohn**: Project administration, Conceptualization, Visualization, Supervision, Validation, Writing - Review & Editing.

## Acknowledgments

This study was supported by the National Research Foundation of Korea grant funded by the Korea government (MSIT) [NRF-2022R1A5A1030054; NRF-RS-2024-00345498; NRF-RS-2023-00281508]; and by Korea Institute for Advancement of Technology (KIAT) grant funded by the Korea Government (MOTIE) [HRD Program for Industrial Innovation - P0023676]. Y.I. is funded by the Deutsche Forschungsgemeinschaft (DFG, German Research Foundation) – 519607530. B.G., F.K., and Y.I. acknowledge funding from the European Research Council (ERC) under the European Union's Horizon 2020 research and innovation program (Grant Agreement No. 865855).

## Declaration of Competing Interest

The authors declare that they have no known competing financial interests or personal relationships that could have appeared to influence the work reported in this study.
34

**Table 1**

Chemical compositions (at.%) obtained via thermodynamic calculation for FCC, HCP+D0$_{19}$, and L1$_2$ phases in the (CoNi)$_{85}$Mo$_{12}$Al$_3$ (3Al-A) and (CoNi)$_{81}$Mo$_{12}$Al$_7$ (7Al-A) alloys aged at 700 °C.

| Alloy | Phase | Co | Ni | Mo | Al |
|---|---|---|---|---|---|
| (CoNi)$_{85}$Mo$_{12}$Al$_3$ | FCC | 39.3 | 49.2 | 7.2 | 4.3 |
| | HCP | 49.6 | 27.7 | 22.7 | < 0.1 |
| (CoNi)$_{81}$Mo$_{12}$Al$_7$ | FCC | 42.6 | 44.6 | 6.5 | 6.2 |
| | L1$_2$ | 18.3 | 59.6 | 4.1 | 18.1 |
| | HCP | 54.0 | 23.6 | 22.4 | < 0.1 |

**Table 2**

Stable and unstable SFEs (mJ/m$^2$) of the L1$_2$ phases obtained by DFT simulations.

| L1$_2$ composition | $\gamma_{UCSF}$ | $\gamma_{CSF}$ | $\gamma_{UAPB}$ | $\gamma_{APB}$ | $\gamma_{USISF}$ | $\gamma_{SISF}$ |
|---|---|---|---|---|---|---|
| Ni$_3$Al | 267 | 192 | 569 | 168 | 259 | 74 |
| (Co$_{1/3}$Ni$_{2/3}$)$_3$Al | 237 | 132 | 498 | 144 | 266 | 67 |
| (Co$_{1/3}$Ni$_{2/3}$)$_3$(Al$_{3/4}$Mo$_{1/4}$) | 344 | 284 | 649 | 288 | 422 | 136 |
| (Co$_{1/3}$Ni$_{2/3}$)$_3$(Al$_{1/2}$Mo$_{1/2}$) | 436 | 381 | 715 | 329 | 454 | 30 |

**Table 3**

Room-temperature tensile properties and average grain sizes for the as-annealed and aged $(CoNi)_{88-x}Mo_{12}Al_x$ alloys.

| Alloy | Grain size (μm) | YS (MPa) | TS (MPa) | El. (%) |
|---|---|---|---|---|
| 3Al | 3.5 ± 1.9 | 652 ± 4 | 1088 ± 9 | 59 ± 2 |
| 3Al-A | 3.9 ± 2.5 | 764 ± 1 | 1202 ± 8 | 48 ± 1 |
| 7Al | 3.6 ± 1.5 | 668 ± 13 | 1127 ± 7 | 62 ± 1 |
| 7Al-A | 4.2 ± 2.2 | 1086 ± 1 | 1520 ± 6 | 35 ± 0.4 |

**Figure captions**

**Fig. 1.** (a) Calculated phase diagram of the $(CoNi)_{88-x}Mo_{12}Al_x$ system as a function of the Al content. Equilibrium phase fractions as a function of temperature for the (b) $(CoNi)_{85}Mo_{12}Al_3$ and (c) $(CoNi)_{81}Mo_{12}Al_7$ alloys.

**Fig. 2.** (a) XRD profiles and (b–e) EBSD IPF maps for the 3Al and 7Al alloys annealed at 1000 °C for 2 min, and the 3Al-A and 7Al-A alloys aged at 700 °C for 24 h. Red, black, and blue symbols in (a) indicate the FCC, HCP, and $D0_{19}$ phases, respectively.

**Fig. 3.** SEM-BSE images of the (a) 3Al, (b) 3Al-A, (c) 7Al, and (d) 7Al-A alloys, showing FCC matrix with annealing twins and plate-like precipitates along grain boundaries and triple junctions.

**Fig. 4**. (a) TEM-BF and (b,c) corresponding DF images of grain-boundary precipitates in the 3Al-A alloy with SADP patterns for the (b) HCP + $D0_{19}$ layers and (c) FCC layers. (d) HR-TEM images of the white box in (c) showing the FCC layers and HCP + $D0_{19}$ layers, with FFT patterns of each phase. TEM-EDS line profile along the orange arrow of (e) and (f) shown in (b) and (c), respectively.

**Fig. 5.** (a) STEM image with the corresponding SADPs showing FCC matrix, $L1_2$ precipitate, and HCP + $D0_{19}$ precipitate in 7Al-A alloy. DF images of ($b_1$) $L1_2$ precipitates near the grain boundary and ($b_2$) $L1_2$ precipitates in FCC matrix, corresponding to the regions indicated in (a). (c) DF image of HCP + $D0_{19}$ precipitate in (a). (d) TEM-EDS line profiles along the orange arrow shown in (a).

**Fig. 6.** APT elemental distribution analysis of the 3Al-A and 7Al-A alloys. (a) Three-dimensional reconstruction map for the 3Al-A alloy and (b) one-dimensional compositional profiles of each element along the arrow shown in (a). (c) Three-dimensional map for the 7Al-A alloy and (d,e) one-dimensional compositional profiles of each element along the arrow shown in (c).

**Fig. 7.** GSFE curves of (a) the $L1_2$-$Ni_3Al$ and (b) the $L1_2$-$(Co_{1/3}Ni_{2/3})_3(Al_{1/2}Mo_{1/2})$ along the NF-CSF-APB-SISF path.

**Fig. 8.** (a) Room-temperature engineering stress-strain curves of all alloys, (b) strain-hardening rate curves for the 3Al and 3Al-A alloys, and (c) strain-hardening rate curves for the 7Al and 7Al-A alloys.

**Fig. 9.** Deformed microstructure of the 3Al-A and 7Al-A alloys deformed to 5% true strain. STEM images of ($a_1$, $a_2$) 3Al-A and ($b_1$, $b_2$) 7Al-A alloys showing dislocations and SFs developed along {111} planes. HR-TEM image and FFT pattern of the white box region of ($a_3$) 3Al-A and ($b_3$) 7Al-A alloys, showing twin spots and SFs. STEM image of ($a_4$) 3Al-A and ($b_4$) 7Al-A alloy, exhibiting dislocation pile up at the interface between FCC matrix and HCP+$D0_{19}$ precipitate.

**Fig. 10.** Deformed microstructure of the 3Al-A and 7Al-A alloys deformed to 20% true strain. STEM image of ($a_1$) 3Al-A and ($b_1$) 7Al-A alloy, showing well-developed dislocations and SFs network. HR-TEM image and FFT pattern of white box region of ($a_2$) 3Al-A and ($b_2$) 7Al-A alloys showing atomic image and nano-twin formations. STEM images of ($a_3$) 3Al-A and ($b_3$) 7Al-A alloys, showing dislocation pile up at the interface between FCC matrix and HCP + $D0_{19}$ precipitate.

**Fig. 11.** Dislocation dissociation analysis of 3Al-A and 7Al-A alloys deformed to 5% true strain. Weak beam TEM-DF images of ($a_1$) 3Al-A and ($b_1$) 7Al-A alloys using $g = \bar{2}20$, $g = \bar{2}02$, and $g = 0\bar{2}2$, respectively. Schematic diagram of dislocation dissociation and each type of planar fault in ($a_2$) 3Al-A and ($b_2$) 7Al-A alloys.

**Fig. 12.** Ashby plot showing the tensile strength versus the elongation for the present alloy in comparison with those of FCC+$L1_2$[4,55–58] M/HEAs, FCC+$L1_2$+B2[59–61] M/HEAs, FCC+$L1_2$+$L2_1$[62,63] M/HEAs, FCC+B2[64–67] M/HEAs, and FCC+σ[68], μ[69–71], $M_{23}C_6$[72,73] M/HEAs.

# Supplemental Materials:
# Ultrastrong and ductile CoNiMoAl medium-entropy alloys enabled by L1$_2$ nanoprecipitate-induced multiple deformation mechanisms

Min Young Sung,[1] Tae Jin Jang,[1] Sang Yoon Song,[1] Gunjick Lee,[1] KenHee Ryou,[2] Sang-Ho Oh,[3] Byeong-Joo Lee,[3] Pyuck-Pa Choi,[2] Jörg Neugebauer,[4] Blazej Grabowski,[5] Fritz Körmann,[4,5,6] Yuji Ikeda,[4,5] Alireza Zargaran,[7] and Seok Su Sohn[1]

[1]*Department of Materials Science and Engineering, Korea University, 02841 Seoul, South Korea*
[2]*Department of Materials Science and Engineering, Korea Advanced Institute of Science and Technology, 34141 Daejeon, South Korea*
[3]*Department of Materials Science and Engineering, Pohang University of Science and Technology, 37673 Pohang, South Korea*
[4]*Max-Planck-Institut für Eisenforschung, Max-Planck-Straße 1, 40237 Düsseldorf, Germany*
[5]*Institute for Materials Science, University of Stuttgart, Pfaffenwaldring 55, 70569 Stuttgart, Germany*
[6]*Interdisciplinary Centre for Advanced Materials Simulation (ICAMS), Ruhr-Universität Bochum, 44801, Germany*
[7]*Graduate Institute of Ferrous Technology, Pohang University of Science and Technology, 37673 Pohang, South Korea*

## I. SUPPORTING EXPERIMENTAL RESULTS

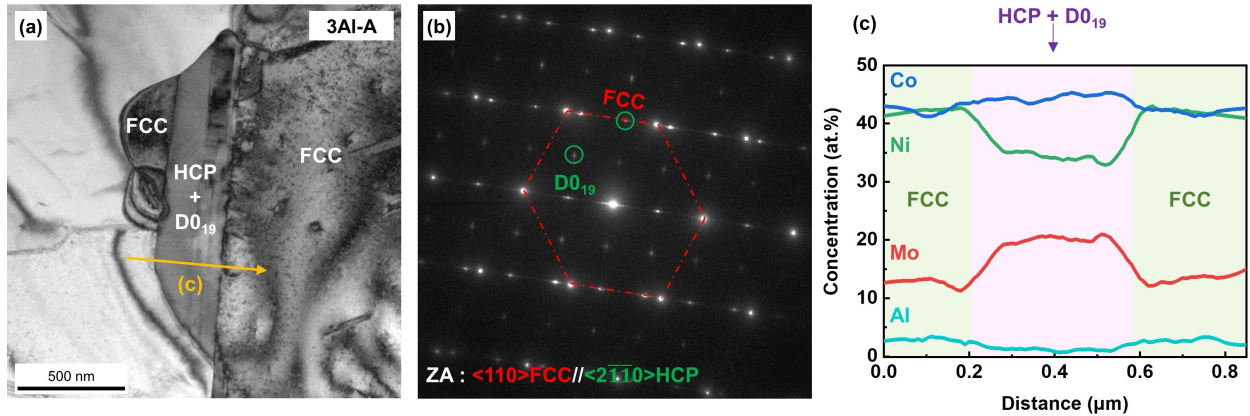

Figure S1. (a) Low magnitude STEM image with FCC matrix and HCP+D0$_{19}$ precipitate in 3Al-A alloy, (b) SADP pattern for (a), (c) TEM-EDS line profile of the orange arrow shown in (a).

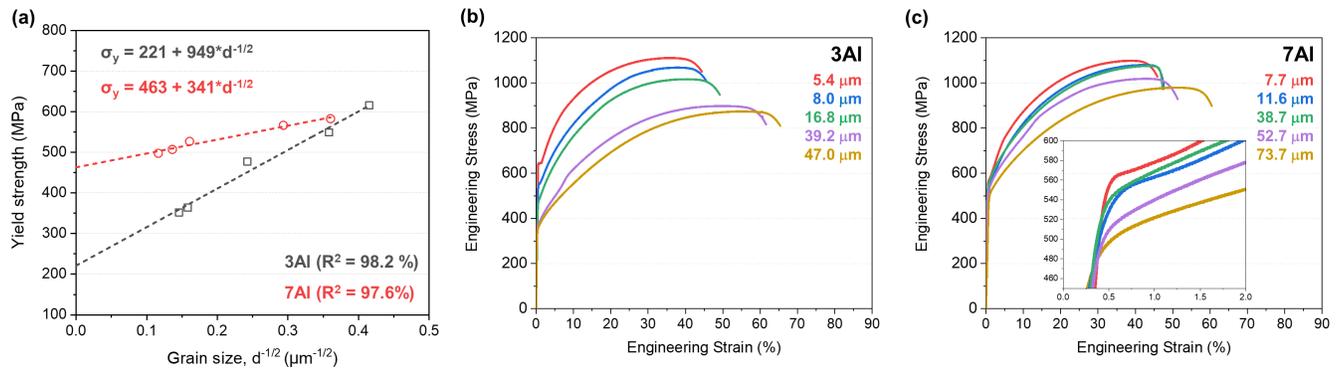

Figure S2. Mechanical properties of the (CoNi)$_{88-x}$Mo$_{12}$Al$_x$ alloys. (a) Yield strength as a function of grain size in the 3Al and 7Al alloys with single FCC phase. Engineering stress–strain curves of the (b) 3Al and (c) 7Al alloys after different heat treatments.

Table S1. Thermal processing, grain size, and tensile properties of the 3Al and 7Al alloys.

| Alloy | Thermal processing | Grain size (μm) | Yield strength (MPa) | Tensile strength (MPa) | Elongation (%) |
|---|---|---|---|---|---|
| 3Al | 1000 °C 10 min | 5.4 ± 3.1 | 628 ± 17 | 1108 ± 4 | 44 ± 5 |
|  | 1000 °C 1 h | 8.0 ± 4.3 | 532 ± 13 | 1059 ± 13 | 46 ± 8 |
|  | 1000 °C 3 h | 16.8 ± 8.2 | 469 ± 11 | 1010 ± 11 | 51 ± 2 |
|  | 1100 °C 30 min | 39.2 ± 19.2 | 369 ± 6 | 905 ± 10 | 61 ± 4 |
|  | 1100 °C 3 h | 47.0 ± 23.8 | 349 ± 5 | 864 ± 14 | 65 ± 6 |
| 7Al | 1000 °C 10 min | 7.7 ± 3.8 | 575 ± 11 | 1104 ± 7 | 46 ± 5 |
|  | 1000 °C 30 min | 11.6 ± 5.7 | 555 ± 8 | 1082 ± 3 | 49 ± 2 |
|  | 1100 °C 10 min | 38.7 ± 16.2 | 536 ± 13 | 1063 ± 11 | 48 ± 6 |
|  | 1100 °C 30 min | 52.7 ± 23.7 | 522 ± 10 | 1027 ± 9 | 54 ± 3 |
|  | 1100 °C 1 h | 73.7 ± 32.5 | 487 ± 7 | 963 ± 12 | 62 ± 1 |

## II. COMPUTATIONAL DETAILS

### A. Models

The following disordered alloys were modeled in the *ab initio* simulations in the present study to study mainly the impact of Mo and Al contents:

- FCC
  - $Co_{48}Ni_{48}$
  - $Co_{42}Ni_{42}Mo_{12}$
  - $Co_{38}Ni_{38}Mo_{12}Al_8$
- $L1_2$
  - $Ni_3Al$
  - $(Co_{1/3}Ni_{2/3})_3Al$
  - $(Co_{1/3}Ni_{2/3})_3(Al_{3/4}Mo_{1/4})$
  - $(Co_{1/3}Ni_{2/3})_3(Al_{1/2}Mo_{1/2})$

These alloys were modeled using 96-atom supercells containing six {111} planes along the *z*-axis. Each plane accommodates 16 atoms and has a dimension of $4d \times 4d$, where $d$ is the first nearest-neighbor atomic distance. The lattice parameters were fixed to 0.35718 nm for $L1_2$ $Ni_3Al$, an experimental value at 25 °C [S1], and to 0.3580 nm for the other alloys, the experimental value of the $L1_2$ phase in the present study. The chemical disorder was modeled based on special quasirandom structure (SQS) [S2] configurations obtained using the ICET code [S3] with the simulated annealing approach. Four SQS configurations were considered for each alloy. Correlation functions of the first, the second, and the third nearest-neighbor doublet clusters and the first nearest-neighbor triplet and quartet clusters were optimized to be close to the ideal values of fully random configurations. For the $L1_2$ phase, configurations of both the Co–Ni and the Mo–Al sublattices were optimized. Figure S3(a) shows one of the thus constructed supercells without stacking faults for $L1_2$ $(Co_{1/3}Ni_{2/3})_3Al$ as an example.

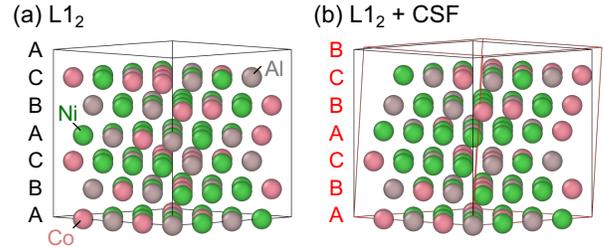

Figure S3. Example for supercell models of chemically disordered $L1_2$ $(Co_{1/3}Ni_{2/3})_3Al$ alloy. (a) Without stacking faults. (b) With a CSF obtained by the tilted-supercell approach. The tilted cell is shown in red, while the original cell is shown in black. The labels A, B, and C correspond to the stacking sequence of the {111} planes. A CSF is introduced at the top of the simulation cell in this example. Visualization was performed using OVITO [S4].

### B. Electronic-structure calculations

*Ab initio* density functional theory (DFT) calculations were performed using the VASP code [S5–S7] with the plane-wave projector augmented wave (PAW) method [S8] and the exchange–correlation functional within the generalized gradient approximation (GGA) of the Perdew–Burke–Ernzerhof (PBE) form [S9]. The plane-wave cutoff energy was set to 400 eV. Reciprocal spaces were sampled by a Γ-centered $4 \times 4 \times 3$ *k*-point mesh for the 96-atom models and the Methfessel–Paxton method [S10] with the smearing width of 0.1 eV. The 3d4s orbitals of Co and Ni, the 4d5s orbitals of Mo, and the 3s3p orbitals of Al were treated as valence states. Total energies were minimized until they converged within $1 \times 10^{-4}$ eV per simulation cell for each ionic step. All calculations were performed considering spin polarization. The cell shape was fixed, and internal atomic positions were relaxed under some constraints as detailed in Sec. II C until all forces on atoms converged within $5 \times 10^{-2}$ eV/Å. The atomic positions for the structures without stacking faults were relaxed from the ideal close-packed positions.

## C. Generalized stacking-fault energies

Generalized stacking-fault energies (GSFEs) are relative energies per fault area when shifting one side of a system with respect to the other side. In general, the GSFEs on a certain lattice plane can be written as a function of the shift vector $\mathbf{u}$ as

$$\gamma(\mathbf{u}) = \frac{E(\mathbf{u}) - E(\mathbf{0})}{A}, \tag{S1}$$

where $E(\mathbf{u})$ is the energy corresponding to $\mathbf{u}$, and $A$ is the fault area. Particularly, $\mathbf{u} = \mathbf{0}$ corresponds to the no-fault (NF) state, and $\gamma(\mathbf{0}) = 0$.

The GSFEs of the investigated alloys for the {111} plane were computed by explicitly calculating several selected points on the GSFE surface followed by a Fourier interpolation. For the explicit calculations, the tilted-supercell approach under the periodic boundary conditions (see, e.g., Appendix A in Gholizadeh et al. [S11] and Sec. II C in Zhang et al. [S12]) was employed. In this approach, the fault is introduced in the simulation cell by tilting the $\langle 111 \rangle$ lattice vector by the shift vector. The ABCABC stacking of the {111} planes is then modified, e.g., to ABC|BCA, where the vertical line indicates the position of the fault plane. The fault can be introduced at any interlayer spaces of {111} planes by shifting the {111} planes above the fault together with the cell tilting. There are thus six possible fault positions in each of our present six-layer models. Figure S3(b) shows one of the thus created models as an example.

Let $\mathbf{a}_i$ ($i \in \{1, 2, 3\}$) be the basis vectors of the conventional cubic FCC and the L1$_2$ lattices and $\mathbf{a}_{\langle uvw \rangle} = u\mathbf{a}_1 + v\mathbf{a}_2 + w\mathbf{a}_3$ ($u, v, w \in \mathbb{Z}$). In the present study, the shift vector on the {111} plane was given by

$$\mathbf{u} = x_1 \mathbf{a}_{\langle \bar{1}2\bar{1} \rangle} + x_2 \mathbf{a}_{\langle \bar{2}11 \rangle} \quad (x_1, x_2 \in \mathbb{R}). \tag{S2}$$

The explicit calculations were carried out on the grid points

$$x_1 = \frac{n_1}{N}, \quad x_2 = \frac{n_2}{N}, \tag{S3}$$

where $n_1, n_2 \in \mathbb{Z}$ and $N = 12$. Figure S4 visualizes the grid points for both the FCC and the L1$_2$ phases. Note that, in principle, a different set of lattice vectors can also be taken as the references of the shift vector $\mathbf{u}$. For example, $\mathbf{a}_{\langle 01\bar{1} \rangle}$ and $\mathbf{a}_{\langle \bar{1}01 \rangle}$ may be more intuitive due to the similarity of the conventional choice of the lattice basis vectors for the hexagonal lattice. For the analysis on the GSFE surface, however, it may be beneficial to consider the grid points specified in Eqs. (S2) and (S3) with $\mathbf{a}_{\langle \bar{1}2\bar{1} \rangle}$ and $\mathbf{a}_{\langle \bar{2}11 \rangle}$ as the references because these two vectors are along the dislocation-dissociation directions and thus because the grid points include the approximate local minima and maxima of the GSFE surface in the hard-sphere model.

There are rotational, reflectional, and translational symmetries for an ideal ordered structure without chemical disorder, resulting in sets of symmetry equivalent points providing the same GSFE. When $N = 12$, there are 6 and 15 such irreducible unique points for the FCC and the L1$_2$ phases, respectively. In Fig. S4, the symmetrically equivalent points are labeled with the same index. For a disordered alloy model, the ideal symmetry is broken. Therefore, the points in each set should generally show different energies. The energy differences originate from the fluctuation of the local chemical composition near the fault [S13, S14] and should therefore become smaller when the number of supercell models or the horizontal expansion of the supercell models increases. In the present study, in order to apply the FCC and the L1$_2$ symmetry, GSFEs were averaged over symmetrically equivalent points. For the averaging, the shift vectors with the smallest amplitudes were considered, which reside in the yellow-shaded hexagonal regions in Fig. S4 including the edges. Taking the ISF for the FCC phase as an example, for each fault plane, three shifting vectors $\mathbf{u}$ were considered, each of which is along $\langle 2\bar{1}\bar{1} \rangle$, $\langle \bar{1}2\bar{1} \rangle$, and $\langle \bar{1}\bar{1}2 \rangle$. Since four SQS models were considered for each FCC alloy model and since six inter-layer positions to introduce the faults are there in each SQS model, a total of 72 configurations were considered to compute the average energy of the ISF. For each disordered FCC and L1$_2$ phase, 436 and 1444 energies were involved to construct the whole GSFE surface.

Several points on the GSFE surfaces correspond to local minima, namely the ones corresponding to the intrinsic stacking fault (ISF) for the FCC phase and the ones corresponding to the complex stacking fault (CSF), the superlattice intrinsic stacking fault (SISF), and the anti-phase boundary (APB) for the L1$_2$ phase. These local minima should be on top of or near the positions of the hard-sphere model given in Tables S2 and S3. Note that, for the CSF and the APB, the positions of the hard-sphere model are just approximations because the symmetry does not fully constrain the positions of these local minima. The real local minima should therefore be found from an interpolation of the GSFE surfaces.

Figure S5 shows the relaxation procedure of the internal atomic positions in the GSFE calculations. The NF states were first optimized, allowing relaxation of all internal coordinates. The fault planes were then introduced using the tilted-supercell above approach. Finally, for the shift vectors corresponding to the aforementioned local minima, internal atomic positions were fully relaxed. For the other shift vectors, internal atomic positions were relaxed only along the $z$-direction. Note that the source code of VASP was modified to perform this partial relaxation for the selected Cartesian coordinates. In both cases, the cell shape was fixed for the sake of computational cost.

Once the energies at the points for explicit calculations were obtained, the GSFE surface was interpolated based on a two-dimensional discrete Fourier transform. First, since the GSFE surface is periodic along both $\mathbf{a}_{\langle \bar{1}2\bar{1} \rangle}$ and $\mathbf{a}_{\langle \bar{1}2\bar{1} \rangle}$, the Fourier coefficients $A_{k_1 k_2}$ can be written formally as

$$A_{k_1 k_2} = \sum_{n_1=0}^{N'-1} \sum_{n_2=0}^{N'-1} \gamma_{n_1 n_2} \exp\left\{-2\pi \mathrm{i} \left(\frac{k_1 n_1}{N'} + \frac{k_2 x_2}{N'}\right)\right\}, \tag{S4}$$

where $\gamma_{n_1 n_2}$ is the GSFE for the shift vector specified by $n_1$

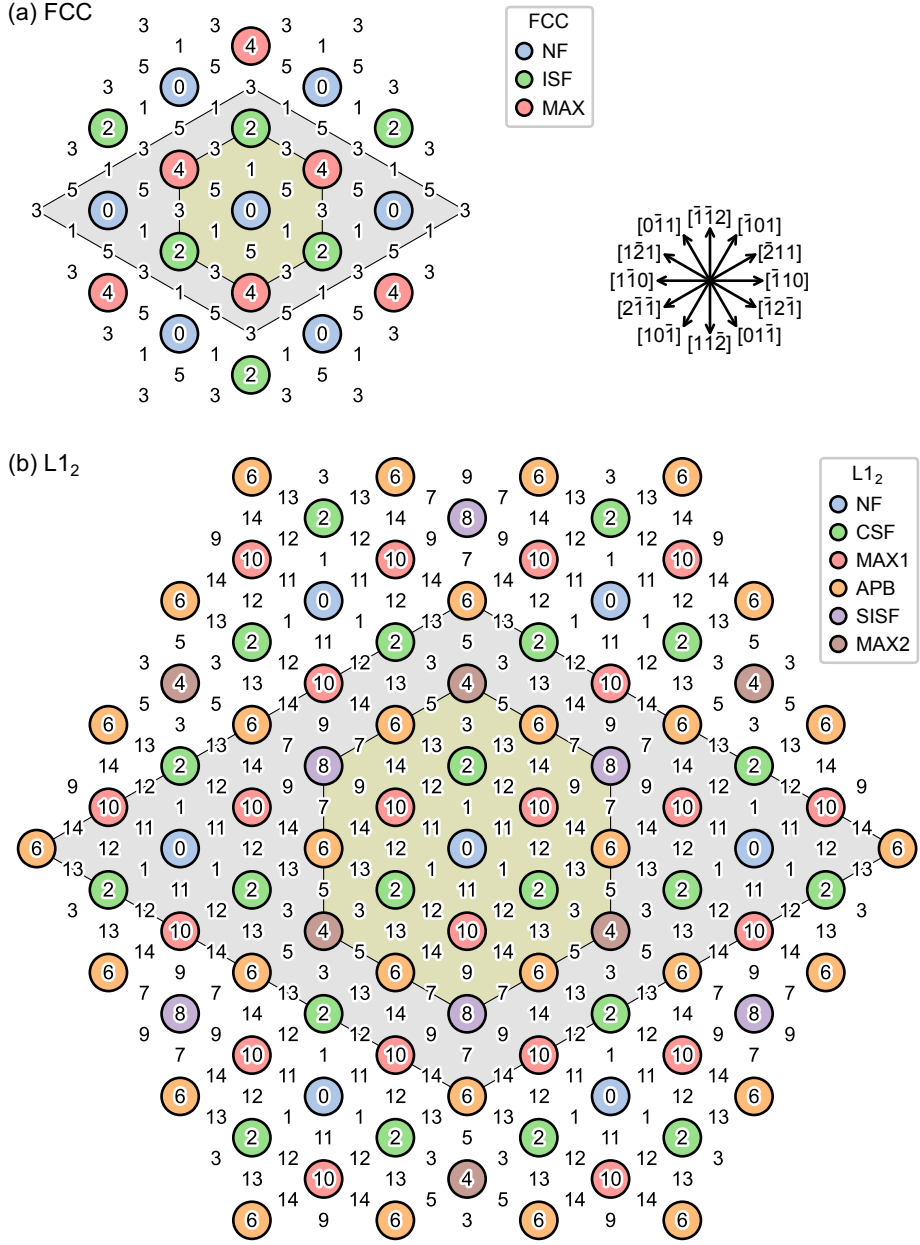

Figure S4. Points on the GSFE surfaces explicitly computed in the present study. Symmetrically equivalent points are labeled with the same number. The yellow-shaded regions are periodic units, while the gray-shaded regions are given in Eq. (S2) with $x_1, x_2 \in [-1/4, +1/4]$ and $x_1, x_2 \in [-1/2, +1/2]$ for the FCC and the L1$_2$ structures, respectively. The periodic unit is twice larger in L1$_2$ than in FCC along each direction.

and $n_2$ in Eq. (S3), $N'$ is $N/2$ and $N$ for the FCC and for the L1$_2$ structures, respectively, and $k_1, k_2 \in \{0, \ldots, N'-1\}$. Note that the grid points reside in the gray-shaded regions in Fig. S3, which contain not one but three periodic units. The GSFEs $\gamma_{n_1 n_2}$ should therefore be provided considering the corresponding translational symmetry. The Fourier coefficients $A_{k_1 k_2}$ were computed using the fast Fourier transform (FFT). Based on the thus obtained Fourier coefficients, the interpolated and GSFE surfaces are given by

$$\gamma(x_1, x_2) = \frac{1}{N'^2} \sum_{k_1'} \sum_{k_2'} c_{k_1'} c_{k_2'} A_{k_1' k_2'} \exp(+2\pi i (k_1' x_1 + k_2' x_2)), \quad (S5)$$

where $A_{k_1' k_2'} = A_{k_1 k_2}$ with $k_i = k_i' \mod N'$. Here, in order to minimize the oscillation of the interpolants, the

Table S2. Representative points on the GSFE surfaces of the FCC phases explicitly computed in the present study. The coordinates are specified by $n_1$ and $n_2$ in Eq. (S3) for $N = 12$. The indices and the labels of the positions correspond to those in Fig. S4(a). The points with bold labels are on top of or near the corresponding local minima. $N_{eq}$ is the number of symmetrically equivalent points in yellow-shaded region in Fig. S4(a) including edges. $N_{calc}$ is the total number of calculations for the disordered alloys to obtain the mean value. The thus obtained GSFEs (mJ/m$^2$) with the standard errors of the means are also shown for the investigated FCC phases.

| | Label | $n_1$ | $n_2$ | $N_{eq}$ | $N_{calc}$ | $Co_{48}Ni_{48}$ | $Co_{42}Ni_{42}Mo_{12}$ | $Co_{38}Ni_{38}Mo_{12}Al_8$ |
|---|---|---|---|---|---|---|---|---|
| 0 | **NF** | 0 | 0 | 1 | 4 | 0 | 0 | 0 |
| 1 | — | 1 | 0 | 3 | 72 | 253 ± 1 | 260 ± 3 | 268 ± 5 |
| 2 | **ISF** | 2 | 0 | 3 | 72 | 8 ± 3 | −24 ± 4 | −7 ± 8 |
| 3 | — | 1 | 1 | 6 | 144 | 698 ± 1 | 752 ± 3 | 798 ± 6 |
| 4 | MAX | 0 | 2 | 3 | 72 | 1193 ± 3 | 1272 ± 9 | 1377 ± 12 |
| 5 | — | 0 | 1 | 3 | 72 | 676 ± 1 | 736 ± 4 | 774 ± 6 |

Table S3. Representative points on the GSFE surfaces of the L1$_2$ phases explicitly computed in the present study. The coordinates are specified by $n_1$ and $n_2$ in Eq. (S3) for $N = 12$. The indices and the labels of the positions correspond to those in Fig. S4(b). The points with bold labels are on top of or near the corresponding local minima. $N_{eq}$ is the number of symmetrically equivalent points in yellow-shaded region in Fig. S4(b) including edges. $N_{calc}$ is the total number of calculations for the disordered alloys to obtain the mean value. The thus obtained GSFEs (mJ/m$^2$) with the standard errors of the means are also shown for the investigated L1$_2$ phases.

| | Label | $n_1$ | $n_2$ | $N_{eq}$ | $N_{calc}$ | $Ni_3Al$ | $(Co_{1/3}Ni_{2/3})_3Al$ | $(Co_{1/3}Ni_{2/3})_3(Al_{3/4}Mo_{1/4})$ | $(Co_{1/3}Ni_{2/3})_3(Al_{1/2}Mo_{1/2})$ |
|---|---|---|---|---|---|---|---|---|---|
| 0 | **NF** | 0 | 0 | 1 | 4 | 0 | 0 | 0 | 0 |
| 1 | — | 1 | 0 | 3 | 72 | 263 | 236 ± 1 | 329 ± 6 | 404 ± 4 |
| 2 | **CSF** | 2 | 0 | 3 | 72 | 226 | 160 ± 2 | 312 ± 8 | 400 ± 5 |
| 3 | — | 3 | 0 | 3 | 72 | 1308 | 1211 ± 4 | 1327 ± 5 | 1363 ± 6 |
| 4 | MAX2 | 4 | 0 | 3 | 72 | 2017 | 1864 ± 3 | 1989 ± 5 | 2040 ± 7 |
| 5 | — | 3 | 1 | 6 | 144 | 1312 | 1211 ± 2 | 1317 ± 4 | 1342 ± 4 |
| 6 | **APB** | 2 | 2 | 6 | 144 | 210 | 172 ± 2 | 306 ± 4 | 340 ± 3 |
| 7 | — | 1 | 3 | 6 | 144 | 259 | 266 ± 1 | 422 ± 4 | 443 ± 3 |
| 8 | **SISF** | 0 | 4 | 3 | 72 | 74 | 67 ± 1 | 136 ± 3 | 30 ± 6 |
| 9 | — | 0 | 3 | 3 | 72 | 768 | 757 ± 4 | 952 ± 8 | 980 ± 6 |
| 10 | MAX1 | 0 | 2 | 3 | 72 | 1444 | 1408 ± 2 | 1567 ± 5 | 1618 ± 5 |
| 11 | — | 0 | 1 | 3 | 72 | 767 | 726 ± 1 | 832 ± 5 | 908 ± 4 |
| 12 | — | 1 | 1 | 6 | 144 | 851 | 812 ± 2 | 950 ± 4 | 1027 ± 4 |
| 13 | — | 2 | 1 | 6 | 144 | 569 | 498 ± 2 | 649 ± 5 | 714 ± 3 |
| 14 | — | 1 | 2 | 6 | 144 | 823 | 809 ± 2 | 965 ± 5 | 1029 ± 4 |

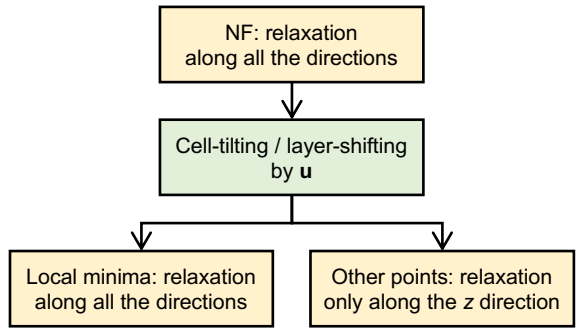

Figure S5. Relaxation procedure of internal atomic positions in the GSFE calculations.

wavenumbers are shifted as

$$k'_1, k'_2 \in \begin{cases} \{-N'/2, \ldots, N'/2\} & \text{if } N' \text{ is even} \\ \{-(N'-1)/2, \ldots, (N'-1)/2\} & \text{if } N' \text{ is odd} \end{cases}. \quad (S6)$$

The coefficients $c_{k'_1}$ and $c_{k'_2}$ are to adapt the Nyquist components and given by

$$c_{k'_i} = \begin{cases} 1/2 & \text{if } N' \text{ is even and } |k'_i| = N'/2 \\ 1 & \text{otherwise} \end{cases}. \quad (S7)$$

The interpolated GSFE surface in Eq. (S5) does not yet show the symmetry of the crystal for arbitrary shift vectors; only the points for the explicit DFT calculations show the symmetry. Suppose that there are $N_s$ symmetry operations and that $\tilde{x}_{1,i}$ and $\tilde{x}_{2,i}$ are the coordinates after a rotation given by

$$\begin{pmatrix} \tilde{x}_{1,i} \\ \tilde{x}_{2,i} \end{pmatrix} = \mathbf{R}_i \begin{pmatrix} x_1 \\ x_2 \end{pmatrix}, \quad (S8)$$

where $\mathbf{R}_i$ is the $i$-th rotational matrix. The final symmetrized GSFE surface is obtained as

$$\tilde{\gamma}(x_1, x_2) = \frac{1}{N_s} \sum_{i=1}^{N_s} \gamma(\tilde{x}_{1,i}, \tilde{x}_{2,i}). \quad (S9)$$

There are six symmetry operations for the {111} surfaces of the FCC and the L1$_2$ phases. When the shift vector **u** is

S5

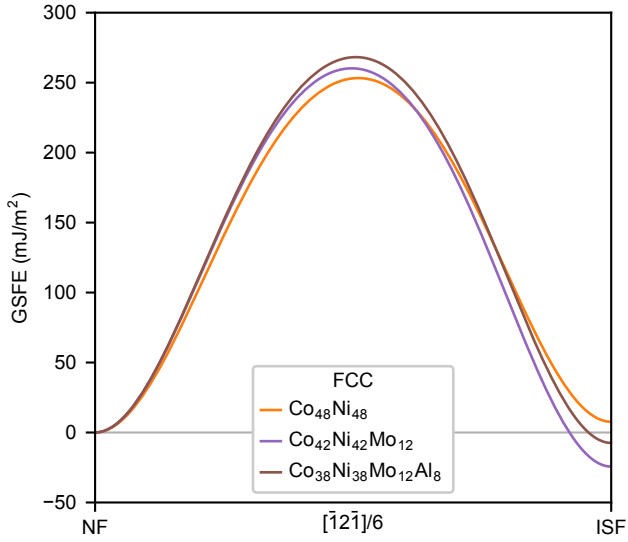

Figure S6. GSFE curves of the A1 phases along the NF–ISF path.

given by $\mathbf{a}_{\langle\bar{1}2\bar{1}\rangle}$ and $\mathbf{a}_{\langle\bar{2}11\rangle}$ in Eq. (S2), the six rotation matrices are given by

$$\mathbf{R}_1 = \begin{pmatrix} +1 & 0 \\ 0 & +1 \end{pmatrix}, \quad \mathbf{R}_2 = \begin{pmatrix} -1 & -1 \\ +1 & 0 \end{pmatrix},$$
$$\mathbf{R}_3 = \begin{pmatrix} 0 & +1 \\ -1 & -1 \end{pmatrix}, \quad \mathbf{R}_4 = \begin{pmatrix} 0 & -1 \\ -1 & 0 \end{pmatrix}, \quad (S10)$$
$$\mathbf{R}_5 = \begin{pmatrix} +1 & +1 \\ 0 & -1 \end{pmatrix}, \quad \mathbf{R}_6 = \begin{pmatrix} -1 & 0 \\ +1 & +1 \end{pmatrix}.$$

To see this, take the positive rotation by 120° along the [111] axis as an example. The above two reference vectors are transformed as

$$\mathbf{a}_{\langle\bar{1}2\bar{1}\rangle} \to -\mathbf{a}_{\langle\bar{1}2\bar{1}\rangle} + \mathbf{a}_{\langle\bar{2}11\rangle}, \quad (S11)$$
$$\mathbf{a}_{\langle\bar{2}11\rangle} \to -\mathbf{a}_{\langle\bar{1}2\bar{1}\rangle}. \quad (S12)$$

The transformed shift vector $\tilde{\mathbf{u}}$ is thus given by

$$\tilde{\mathbf{u}} = x_1(-\mathbf{a}_{\langle\bar{1}2\bar{1}\rangle} + \mathbf{a}_{\langle\bar{2}11\rangle}) + x_2(-\mathbf{a}_{\langle\bar{1}2\bar{1}\rangle}) \quad (S13)$$
$$= (-x_1 - x_2)\mathbf{a}_{\langle\bar{1}2\bar{1}\rangle} + x_1 \mathbf{a}_{\langle\bar{2}11\rangle} \quad (S14)$$
$$= \tilde{x}_{1,2}\mathbf{a}_{\langle\bar{1}2\bar{1}\rangle} + \tilde{x}_{2,2}\mathbf{a}_{\langle\bar{2}11\rangle}, \quad (S15)$$

where the coordinates are

$$\begin{pmatrix} \tilde{x}_{1,2} \\ \tilde{x}_{2,2} \end{pmatrix} = \begin{pmatrix} -1 & -1 \\ +1 & 0 \end{pmatrix}\begin{pmatrix} x_1 \\ x_2 \end{pmatrix} = \mathbf{R}_2 \begin{pmatrix} x_1 \\ x_2 \end{pmatrix}. \quad (S16)$$

Note that the expressions of the rotation matrices are different when we take other lattice vectors like $\mathbf{a}_{\langle 01\bar{1}\rangle}$ and $\mathbf{a}_{\langle\bar{1}01\rangle}$ as references.

Figure S6 shows the GSFE curves of the FCC phases along the NF–ISF path. Figure S7 shows the GSFE surfaces of the FCC phases. Table S4 summarizes the local minima and the saddle points on the GSFE surfaces of the FCC phases.

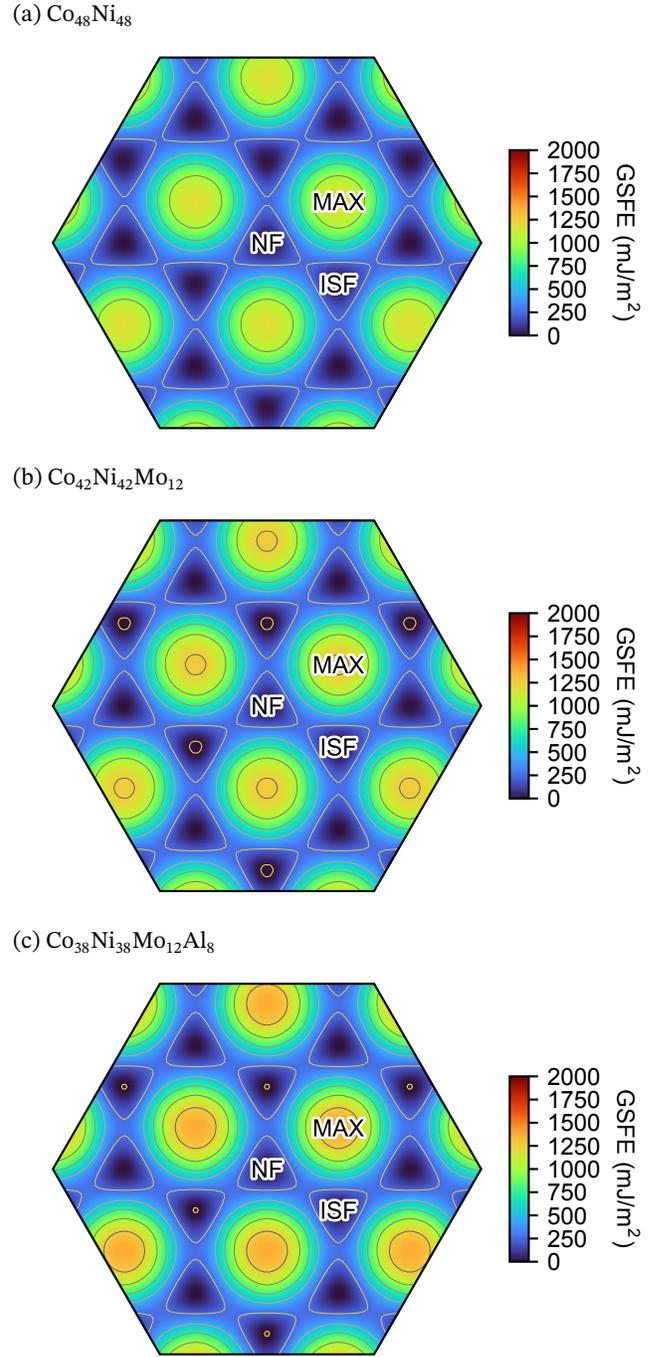

(a) $Co_{48}Ni_{48}$

(b) $Co_{42}Ni_{42}Mo_{12}$

(c) $Co_{38}Ni_{38}Mo_{12}Al_8$

Figure S7. GSFE surface of the FCC phases.

Figure S8 shows the GSFE curves of the L1$_2$ phases along the NF–CSF–APB–SISF path. Figures S9–S12 show the GSFE surfaces of the L1$_2$ phases. Table S5 summarizes the local minima and the saddle points on the GSFE surfaces of the L1$_2$ phases.

Table S4. Local minima and saddle points on the GSFE surfaces (mJ/m$^2$) for the FCC phases. The coordinates are given via Eq. (S2).

|  | Co$_{48}$Ni$_{48}$ | | | Co$_{42}$Ni$_{42}$Mo$_{12}$ | | | Co$_{38}$Ni$_{38}$Mo$_{12}$Al$_8$ | | |
|---|---|---|---|---|---|---|---|---|---|
|  | $x_1$ | $x_2$ | $\gamma$ | $x_1$ | $x_2$ | $\gamma$ | $x_1$ | $x_2$ | $\gamma$ |
| UISF | 0.1700 | 0.0000 | 253 | 0.1658 | 0.0000 | 260 | 0.1684 | 0.0000 | 268 |
| ISF | 0.3333 | 0.0000 | 8 | 0.3333 | 0.0000 | −24 | 0.3333 | 0.0000 | −7 |

Table S5. Local minima and saddle points on the GSFE surfaces (mJ/m$^2$) for the L1$_2$ phases. The coordinates are given via Eq. (S2). "UAPB0" is the approximate position of the unstable APB state constrained on the path between the CSF and the APB states in the hard-sphere model, which is discussed in the main text to facilitate visualization, while "UAPB" is the true saddle point.

|  | Ni$_3$Al | | | (Co$_{1/3}$Ni$_{2/3}$)$_3$Al | | | (Co$_{1/3}$Ni$_{2/3}$)$_3$(Al$_{3/4}$Mo$_{1/4}$) | | | (Co$_{1/3}$Ni$_{2/3}$)$_3$(Al$_{1/2}$Mo$_{1/2}$) | | |
|---|---|---|---|---|---|---|---|---|---|---|---|---|
|  | $x_1$ | $x_2$ | $\gamma$ | $x_1$ | $x_2$ | $\gamma$ | $x_1$ | $x_2$ | $\gamma$ | $x_1$ | $x_2$ | $\gamma$ |
| UCSF | 0.0926 | 0.0000 | 267 | 0.0879 | 0.0000 | 237 | 0.1004 | 0.0000 | 344 | 0.1068 | 0.0000 | 436 |
| CSF | 0.1489 | 0.0000 | 192 | 0.1511 | 0.0000 | 132 | 0.1504 | 0.0000 | 284 | 0.1531 | 0.0000 | 381 |
| UAPB0 | 0.1667 | 0.0827 | 569 | 0.1667 | 0.0838 | 498 | 0.1667 | 0.0821 | 649 | 0.1667 | 0.0797 | 715 |
| UAPB | 0.1523 | 0.0895 | 544 | 0.1549 | 0.0896 | 481 | 0.1549 | 0.0883 | 633 | 0.1558 | 0.0855 | 702 |
| APB | 0.1469 | 0.1864 | 168 | 0.1509 | 0.1824 | 144 | 0.1546 | 0.1787 | 288 | 0.1577 | 0.1757 | 329 |
| USISF | 0.0815 | 0.2519 | 259 | 0.0817 | 0.2516 | 266 | 0.0863 | 0.2471 | 422 | 0.0961 | 0.2372 | 454 |
| SISF | 0.0000 | 0.3333 | 74 | 0.0000 | 0.3333 | 67 | 0.0000 | 0.3333 | 136 | 0.0000 | 0.3333 | 30 |

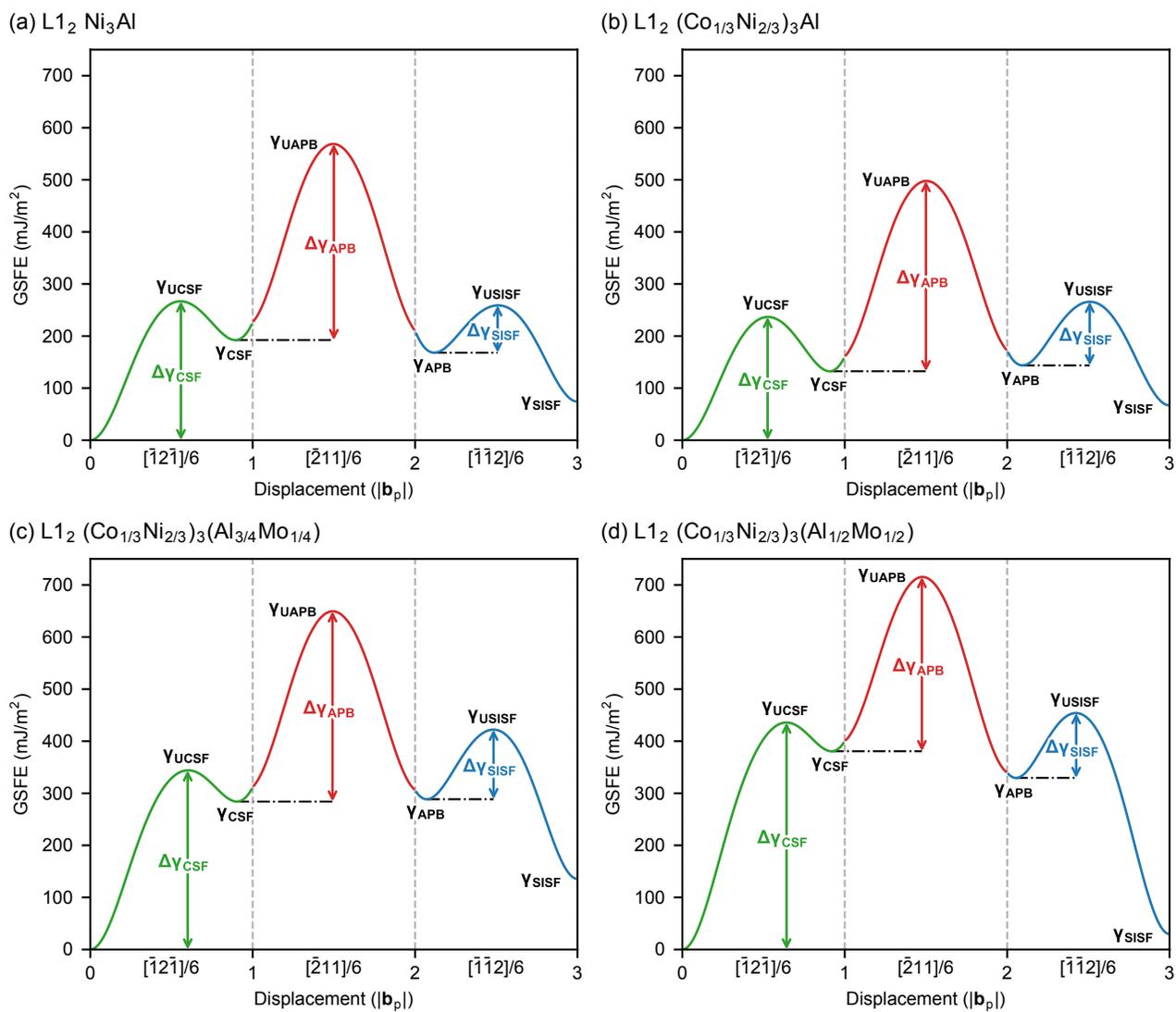

Figure S8. GSFE curves of the L1$_2$ phases along the NF–CSF–APB–SISF path.

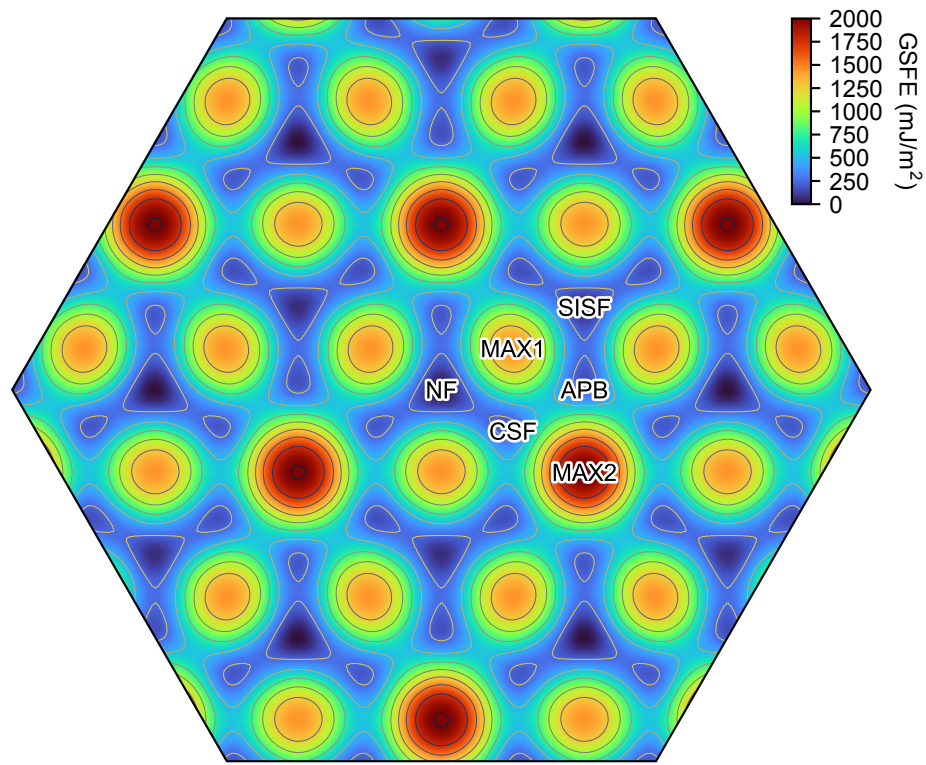

Figure S9. GSFE surface of L1$_2$ Ni$_3$Al.

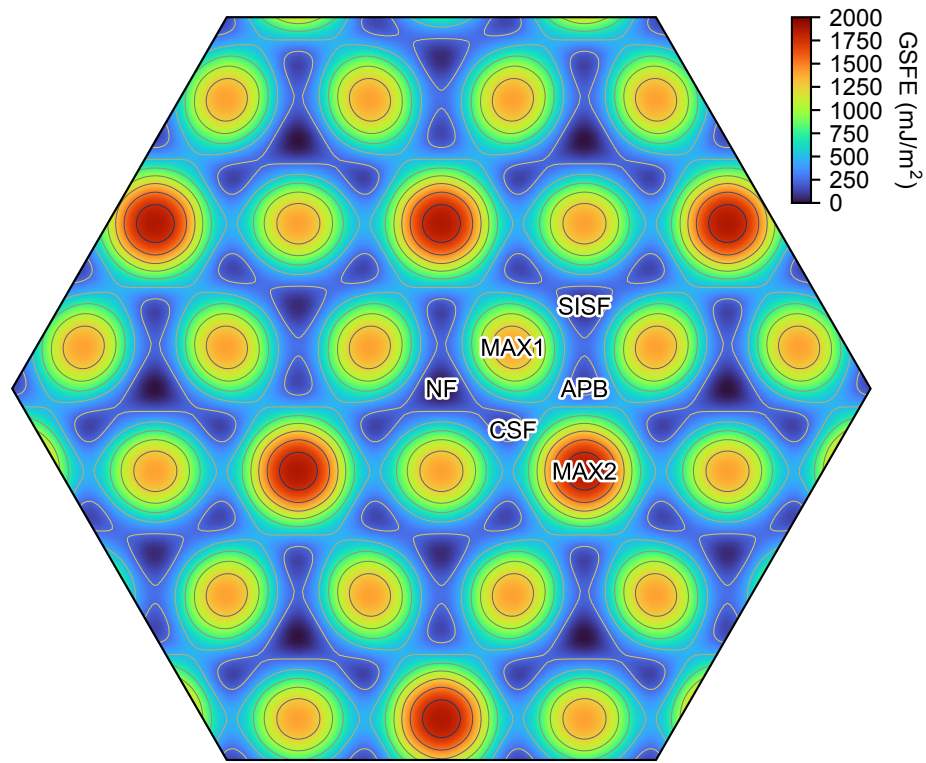

Figure S10. GSFE surface of L1$_2$ (Co$_{2/3}$Ni$_{1/3}$)$_3$Al.

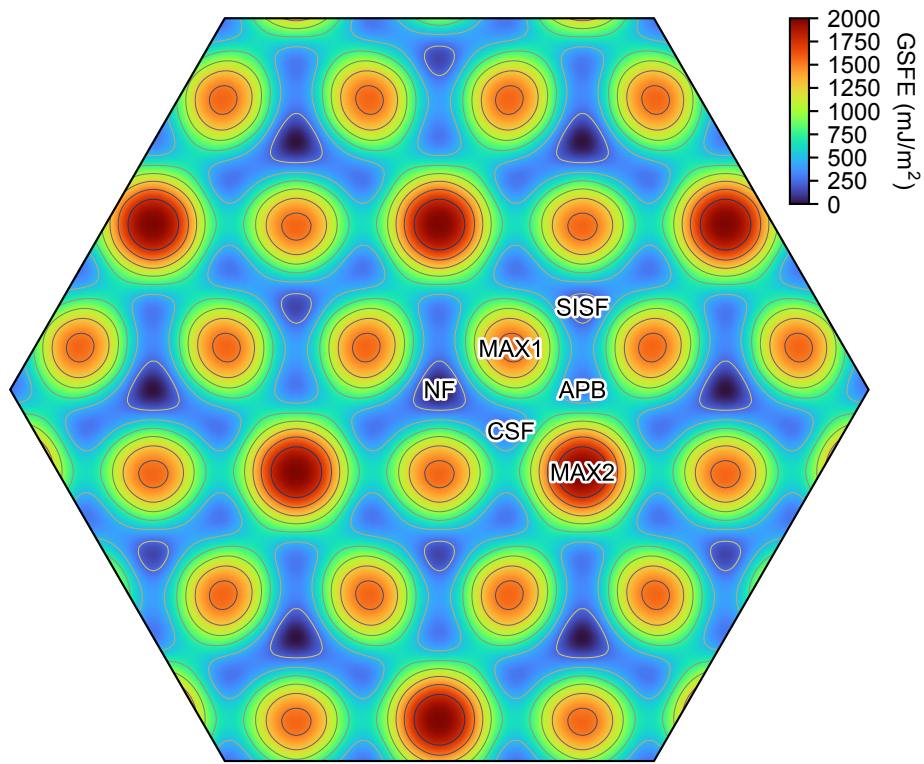

Figure S11. GSFE surface of L1$_2$ (Co$_{2/3}$Ni$_{1/3}$)$_3$(Al$_{3/4}$Mo$_{1/4}$).

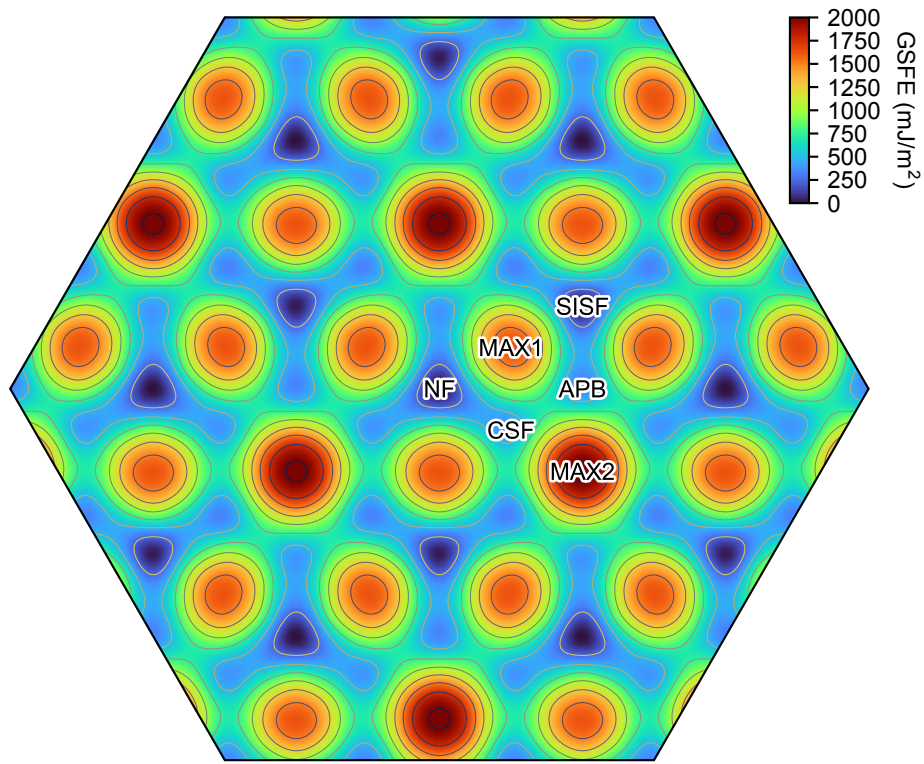

Figure S12. GSFE surface of L1$_2$ (Co$_{2/3}$Ni$_{1/3}$)$_3$(Al$_{1/2}$Mo$_{1/2}$).